\documentclass[aps,prd,twocolumn,superscriptaddress,floatfix,showpacs]{revtex4}
\usepackage{graphics}
\usepackage{graphicx}
\usepackage{epsfig}
\usepackage{amsmath}
\usepackage{amsfonts}
\usepackage{bm}
\usepackage{url}
\usepackage{float}
\usepackage{multirow}

\def\beq{\begin{equation}}
\def\eeq{\end{equation}}
\def\bea{\begin{eqnarray}}
\def\eea{\end{eqnarray}}

\def\dd{\text{d}}

\def\gsim{ \lower .75ex \hbox{$\sim$} \llap{\raise .27ex \hbox{$>$}} }
\def\lsim{ \lower .75ex\hbox{$\sim$} \llap{\raise .27ex \hbox{$<$}} }

\def\m{\mu}
\def\n{\nu}
\def\lm{\lambda^{\m}}
\def\lnu{\lambda^{\nu}}
\def\ldl{\ln D_L}
\def\ci{\cos\iota}

\begin{document}
%% \normalsize  
\input epsf.tex

%\title{A Hamiltonian Monte Carlo for parameter estimation of gravitational waves emitted by binary neutron star coalescences}
\title{Bayesian inference for binary neutron star inspirals using a Hamiltonian Monte Carlo Algorithm}
\author{ Yann \surname{Bouffanais}}
\email[]{bouffana@apc.in2p3.fr}
\author{Edward K. \surname{Porter}}
\email[]{porter@apc.in2p3.fr}
\vspace{1cm}
\affiliation{APC, Universit\'e Paris Diderot,\\ CNRS/IN2P3, CEA/Irfu, Observatoire de Paris, Sorbonne Paris Cit\'e, \\10 rue A. Domon et L. Duquet, 75205 Paris Cedex 13, France}
\vspace{1cm}
\begin{abstract}
The coalescence of binary neutron stars are one of the main sources of gravitational waves for ground-based gravitational wave detectors.  As Bayesian inference for binary neutron stars
is computationally expensive, more efficient and faster converging algorithms are always needed.  In this work, we conduct a feasibility study using a Hamiltonian Monte Carlo algorithm (HMC).
The HMC is a sampling algorithm that takes advantage of gradient information from the geometry of the parameter space to efficiently sample from the posterior distribution, allowing the
algorithm to avoid the random-walk behaviour commonly associated with stochastic samplers.  As well as tuning the algorithm's free parameters specifically for gravitational wave astronomy, we
introduce a method for approximating the gradients of the log-likelihood that reduces the runtime for a $10^6$ trajectory run from ten weeks, using numerical derivatives along the Hamiltonian trajectories, to one day, in the case
of non-spinning neutron stars.  Testing our algorithm against a set of neutron star binaries using a detector network composed of Advanced LIGO and Advanced Virgo at optimal design,  we demonstrate that 
not only is our algorithm more efficient than a standard sampler, but a $10^6$ trajectory HMC produces an effective sample size on the order of $10^4 - 10^5$ statistically independent samples.
%Past studies have further demonstrated that, when properly implemented, this method could be 
% as much as $\mathcal{D}$ times more efficient than a Metropolis-Hastings Monte Carlo algorithm, where $\mathcal{D}$ is the dimensionality of the parameter space.   

%So far, one source has been detected during the second observation run of Advanced LIGO and Advanced Virgo. As the detection rate of these sources is expected to increase in the future, it is essential to have efficient data analysis techniques to be able to treat the data.  In this work, we present an application of a Hamiltonian Monte Carlo algorithm for the parameter estimation of binary neutron star coalescence as measured by a three detector network. We tested our algorithm on a set of 10 sources using the TaylorF2 waveform model. We demonstrated that our algorithm was able to properly sample from the posterior distribution and generate statistically independent samples as much as $10$ times faster than a standard Markov Chain Monte Carlo method.
\end{abstract}

\maketitle
%%%%%%%%%%%%%%%%%%%%%%%%%%%%%%%%%%%%%%%%%%%%%%%%%%%%%%%%%%%%%%%%%%%%%%%%%%%%%%%%
%%%%%%%%%%%%%%%%%%%%%%%%%%%%%
%%%%%%%%%%%%%%%%%%%%%%%%%%%%%%%%%%% Introduction 
%%%%%%%%%%%%%%%%%%%%%%%%%%%%%%%%%%%%%%%%%%%%%%%%%%%%%%%%%%%%
%%%%%%%%%%%%%%%%%%%%%%%%%%%%%%%%%%%%%%%%%%%%%%%%%%%%%%%%%%%%%%%%%%%%%%%%%%%%%%%%
%%%%%%%%%%%%%%%%%%%%%%%%%%%%%

\section{Introduction}

In October 2017, the Advanced LIGO/Advanced Virgo collaboration (LVC) announced the first direct detection of gravitational waves (GWs) from the coalescence of a binary neutron star (BNS) system \cite{BNS_LVC_2017,BNS_2018_1,BNS_2018_2}. As well as being detected by the network of ground-based GW detectors, this event was also detected by a number of electromagnetic observatories, heralding the beginning of  multimessenger astronomy \cite{MultiMessenger_BNS}. This detection of GWs from a BNS merger accompanies the number of binary black hole (BBH) mergers that have also been detected by the LVC \cite{gw150914,gw151226,BBHs_01,gw170104,gw170608,gw170814}.

Bayesian inference, based on matched filtering, is the parameter estimation method of choice within the GW community.  Matched filtering is the 
optimal linear filter for signals buried in noise, and uses phase-matching of a GW template to identify signals in the data.  In recent years, stochastic algorithms based on Markov Chain Monte Carlo (MCMC) methods~\cite{hastings_1970,metropolis_1953} have been shown to be well suited to these problems. The LVC collaboration has developed a library of codes called LALInference \cite{veitch_2014,Berry_2014,Farr_2015lna,Singer:2014qca}, that contains both MCMC and Nested sampling~\cite{skilling_2006} algorithms for parameter estimation. These algorithms belong to a family of random walk stochastic samplers which are known to have 
potentially slow convergence properties.  In this work, we investigate a non-random walk sampler that converges faster than standard 
MCMC samplers.

The Hamiltonian Monte Carlo (HMC) is a method implemented by Duane et al for lattice field simulations \cite{duane_1987}. The main idea behind this algorithm is to suppress the random walk behaviour found in most MCMC methods, by simulating Hamiltonian dynamics in parameter phase-space. Thus, instead of randomly proposing jumps in parameter space, the HMC evolves trajectories in phase-space that are informed by the gradient of the 
target distribution.  The HMC is superior to standard MCMC algorithms as it takes advantage of the background geometry of the parameter space and is able to explore distant points faster than other MCMC methods. If the algorithm is well tuned, the acceptance rate of the HMC algorithm is usually very high and the autocorrelation between adjacent points of the chain is low. Empirically, studies have shown that the HMC is approximately $D$ times more efficient than standard MCMC samplers, where $D$ is the dimensionality of the parameter space \cite{Hajian_2006,Porter_2013}. 

As with all stochastic samplers, there are a number of technical hurdles that have prevented the HMC algorithm from more common usage in Bayesian inference.  Firstly,  the algorithm has a number of 
free parameters that need to be tuned for efficient exploration. Secondly,  the simulation of the Hamiltonian dynamics is {\em the} major computational bottleneck which needs to be 
addressed:  the HMC inverts the likelihood surface, and treats the parameter estimation problem
as a ``gravitational" problem.  While simulating Hamiltonian trajectories in phase space, the $D$-dimensional gradients of the target density have to be calculated at successive points along each trajectory.  If there
is no closed-form solution for the gradients, they have to be calculated numerically.  For GW astronomy, this involves generating multiple waveforms for numerical differencing, resulting in trajectory generation times that are too
slow for efficient application.

In this work, we present an implementation of the HMC algorithm applied to the parameter estimation of BNS sources as observed by a three-detector network. As an initial starting point, we use a version of the HMC algorithm developed for supermassive black hole binary parameter estimation with LISA \cite{Porter_2013}. Our initial investigations revealed that this version of the algorithm was insufficient for our needs, and that it was necessary to adapt this code to make it work in the framework of BNS sources.  The main reason for this is that in the former study, the sources had very high signal-to-noise ratios (SNRs), and as a 
consequence, had unimodal posterior distributions.  Conversely, for BNS sources, a number of parameters have multi-modal posteriors requiring major modifications to the algorithm.

It is difficult to directly
compare the performance of our HMC algorithm with the algorithms in LALinference as that package is highly optimised.  As this is a proof of principle test, we compare the HMC algorithm against a Differential Evolution Markov Chain (DEMC).  For the test 
sources, we use a selection of binaries, covering a large variation in parameter space, from a  previous LVC study~\cite{Singer:2014qca}. 

In Section \ref{section_1}, we outline concepts of Bayesian inference and its application to GW data analysis, as well as defining the TaylorF2 model used to model the coalescence of BNS. In Section \ref{section_2}, we introduce the DEMC and HMC algorithm used for this study. In section \ref{section_3}, we discuss optimising the HMC algorithm for BNS parameter estimation. Finally in Section \ref{section_4}, we present a comparison of results obtained from our HMC and DEMC algorithms. Throughout the paper, we use units of $G=c=1$.

%%%%%%%%%%%%%%%%%%%%%%%%%%%%%%%%%%%%%%%%%%%%%%%%%%%%%%%%%%%%%%%%%%%%%%%%%%

\section{Bayesian inference for gravitational wave parameter estimation}
\label{section_1}

A GW detector measures a time-domain strain, $s(t)$, which is a combination of the noise in the detector $n(t)$, and a potential gravitational wave signal $h(t, \lambda^{\mu})$, defined by a set of parameters $\lambda^{\mu}$, i.e.
\begin{equation}
s(t) = h(t;\lambda^{\mu}) + n(t),
\label{presentation_problem_data_analysis}
\end{equation}
where we assume that the noise is stationary and Gaussian.

\begin{figure}
\epsfig{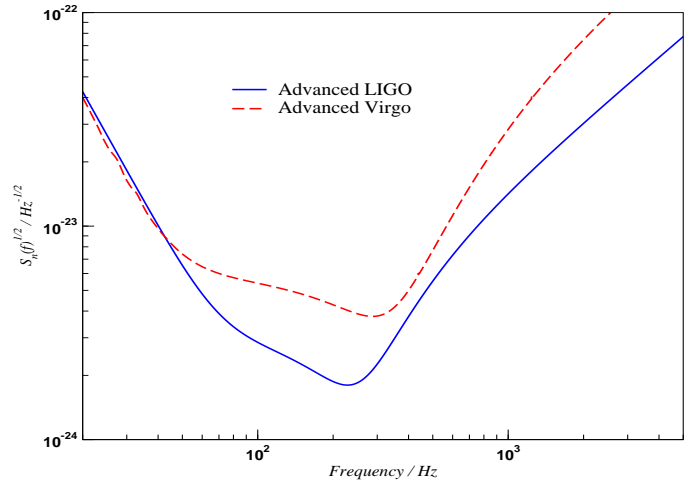}
\caption{Designed sensitivities of advanced LIGO \cite{Sathyaprakash_2009} and advanced Virgo \cite{adv_Virgo_presentation} used in this study.}
\label{designed_sensitivities}
\end{figure}

In GW astronomy, given a data set $s(t)$, a waveform model based on a set of parameters
$\tilde{h}(\lm$), and a noise model for the one-sided power spectral density $S_n(f)$, one can construct the posterior probability distribution $p(\lm |s)$ via Bayes' theorem,
\begin{equation}
p(\lm | s) = \frac{p(s | \lm) \pi(\lm)}{p(s)},
\label{Bayes_theorem}
\end{equation}
where $p(s | \lm) = \mathcal{L}(\lm)$ is the likelihood function (that we will define below), $\pi(\lm)$ is the prior distribution reflecting the a-priori knowledge of our parameters before evaluation of the data and $p(s)$ is the marginalized likelihood or ``evidence" given by,
\begin{equation}
p(s) = \int \mathcal{L}(\lambda^{\mu}) \pi (\lambda^{\mu})  \dd \lambda^{\mu}.
\end{equation}
Under the assumption of Gaussian noise, the likelihood in Eq.~(\ref{Bayes_theorem}) is defined as~\cite{Finn_1992}
\begin{equation}
\mathcal{L}(\lambda^{\mu}) = \exp \left[ - \frac{1}{2} \langle s - h(\lambda^{\mu}) | s - h(\lambda^{\mu})  \right],
\label{likelihood_definition}
\end{equation} 
where the angular brackets denote a noise-weighted inner product,
\begin{equation}
\langle h | g \rangle = 2 \int_{0}^{\infty} \frac{\tilde{h}(f)\tilde{g}^{*}(f) + \tilde{h}^{*}(f)\tilde{g}(f)}{S_{n}(f)} \dd f.
\label{Scalarproduct_GW}
\end{equation}
Here $\tilde{h}(f)$ is the Fourier transform of the time-domain waveform, the asterisk represents a complex conjugate and $S_{n}(f)$ is the one-sided noise power spectral density of the detector. In this study, we consider the design sensitivities of the detectors as represented in Figure \ref{designed_sensitivities}.

As the coalescence of BNS systems take place at high frequencies, where we have low sensitivity, we do not expect to see the merger-ringdown of BNS systems with the current detectors.  As a consequence,
we use the restricted Fourier domain inspiral-only TaylorF2 waveform as our waveform model.  While we expect strong tidal
interactions between the neutron stars as they approach coalescence, we neglect such effects in this work.

The TaylorF2 waveform uses the stationary phase approximation and is written as,
\begin{equation}
\tilde{h}(f) = \sqrt{\frac{5}{24}} \mathcal{M}^{5/6}_{c} \frac{ \mathcal{Q}}{D_{L}\pi^{2/3}} f^{-7/6} \text{e}^{i \Psi(f)},
\label{TaylorF2}
\end{equation}
where $\mathcal{M}_{c}=(m_1 m_2)^{3/5}/(m_1 + m_2)^{1/5}$ is the chirp mass,  $m_1$ and $m_2$ are the individual component masses, $\mathcal{Q}$ is a function depending on the detector response, $D_{L}$ is the luminosity distance and $\Psi(f)$ is the gravitational waveform phase taken at the 3.5 post-Newtonian order (i.e $(v/c)^7$)~\cite{Buonanno_2009}, expressed as 
\begin{equation}
\Psi(f) = 2 \pi f t_{c} - \phi_{c} - \pi / 4 + \frac{3}{128 \eta v^5} \sum_{k=0}^{7} \psi_{k} v^{k},
\label{taylorF2_phase}
\end{equation}
where $\eta=m_1 m_2 / M^2$ is the symmetric mass ratio, $M = m_1 + m_2$ is the total system mass, $t_{c}$ is the coalescence time, $\phi_{c}$ is the phase at coalescence, $v=(\pi M f)^{1/3}$ is the characteristic velocity of the binary and $\psi_{k}$ are the phase coefficients. In this study, we fix $t_{c}$ to be the time it takes for the binary to go from the low frequency cutoff of the detector to the frequency of the last stable circular orbit, i.e. $f_{lso} = 1/(6^{3/2} \pi M)$.  $\mathcal{Q}$ is a function of the location and orientation of the source, and can be written as,
\begin{eqnarray}
\mathcal{Q} &=& \left[ \left(\frac{1}{2}(1 + \cos^{2}(\iota))F^{+}(\alpha, \delta, \psi) \right)^{2}  \right. \nonumber \\
 && +\left. \left(\ci F^{\times}(\alpha, \delta, \psi) \right)^{2} \right]^{1/2},
\end{eqnarray}
where $\ci=\hat{n}.\hat{L}$ is the inclination of the orbital plane, given as the angle between a unit vector along the line of sight, $\hat{n}$, and the 
total angular momentum vector of the binary, $\hat{L}$.    ($\alpha,\delta$) are the right ascension and declination of the source and $\psi$ is the polarisation angle. The functions $F^{+}(\alpha, \delta, \psi)$ and $F^{\times}(\alpha, \delta, \psi)$ are the detector network responses and are specific to each detector~\cite{Anderson_2000}.

As GWs travel with finite speed, they will arrive at the various detectors at different times depending on the sky position of the source. These time delays will induce a change in the phase of the waveform that can be used to infer the position of the source via triangulation. If we consider the three-detector network formed of the two aLIGO and aVirgo detectors,
 and take aLIGO at Hanford as the point of reference, the measured phases at the different detectors are given by,
\begin{eqnarray}
\Psi_{H}(f) &=& \Psi(f),  \\
\Psi_{L}(f) &=& \Psi(f) - 2 \pi f \Delta_{L/H},  \\
\Psi_{V}(f) &=& \Psi(f)- 2 \pi f \Delta_{V/H},  
\end{eqnarray}
where $\Delta_{L/H}$ is the time delay of arrival at Livingston and $\Delta_{V/H}$ is the time delay of arrival at Virgo. Since we have a coherent detection by the network, we can express the $N$-detector network SNR and log-likelihood as,
\begin{eqnarray}
\rho_{net} &=& \sqrt{\sum_{d=1}^{N} \left(\rho_{d}\right)^{2} }, \\
\left( \ln \mathcal{L} \right)_{net} &=&\sum_{d=1}^{N} \left( \ln \mathcal{L} \right)_{d},
\end{eqnarray}
where the single detector SNR is given by
\beq
\rho_d = \frac{\langle s | h \rangle}{\sqrt{\langle h | h \rangle}}.
\eeq

%%%%%%%%%%%%%%%%%%%%%%%%%%%%%%%%%%%%%%%%%%%%%%%%%%%%%%%%%%%%%%%%%%%%

\section{Markov Chain Monte Carlo Methods}
\label{section_2}

Markov Chain Monte Carlo (MCMC) methods are stochastic algorithms that are particularly well suited to Bayesian inference for GW sources~\cite{veitch_2014,Rodriguez_2013,Berry_2014,Farr_2015lna,Singer:2014qca,vanderSluys_2008,porter_2009,Cornish_2006,porter_fisher_2015,Cornish_2005_data}. As stated earlier, this work is a feasibility study to test the application of the HMC algorithm to the parameter estimation of 
BNS sources.  Due to the fact that our algorithm is not optimized, it would be unfair to make a direct comparison with the  optimized codes within 
the LALinference package.  To make as close a comparison
as possible, we decided to test our HMC algorithm against a mixture-model Differential Evolution Markov Chain (DEMC) that has been used in previous GW studies~\cite{porter_fisher_2015,Porter:2015bea}.
%In this section, we present two specific Markov Chain Monte Carlo methods: the Differential Evolution Markov Chain (DEMC) and the Hamiltonian Monte Carlo (HMC) algorithm.  

\subsection{Differential Evolution Markov Chain}
The DEMC is a sampling method that combines two algorithms, namely the Metropolis-Hastings Markov Chain (MHMC)~\cite{metropolis_1953,hastings_1970} and Differential Evolution (DE)~\cite{storn_1997} algorithms.  In this section, we describe the main features of these algorithms and how they are combined to form the DEMC.  

The Metropolis-Hastings variant of the MCMC algorithm works as follows in GW astronomy : starting with a signal $s(t)$, we randomly choose a starting point $x(\lm)$ within a region of the parameter space bounded
by the prior probabilities $\pi(\lm)$ (which we assume to contain the true solution).  We then draw from a transition kernel and propose a jump to another point in parameter space $x'(\lnu)$.  To compare both points, we evaluate the Metropolis-Hastings ratio
\begin{equation}
H = \frac{\pi(x')p(s|x')q(x|x')}{\pi(x)p(s|x)q(x'|x)}.
\end{equation}
Here $q(x|x')$ is a function that defines the jump proposal distribution, and all other quantities are previously defined.  This jump is then accepted with probability $\alpha = min(1,H)$, otherwise the chain stays at $x(\lm)$.   In 
order to improve the overall acceptance rate, the most efficient proposal distribution or jumps in the parameter space is, in general, a multi-variate Gaussian distribution.  However, to ensure that the proper dimensional scalings are taken into account, we can assume that the Fisher 
Information matrix (FIM), defined by
\begin{equation}
\Gamma_{\mu\nu} =  - \mathbb{E} \left[ \frac{\partial^{2} \ln \mathcal{L}}{\partial \lambda^{\mu} \lambda^{\nu}} \right] = \Big\langle \frac{\partial \tilde{h}}{\partial \lambda^{\mu}} \Big\rvert \frac{\partial \tilde{h}}{\partial \lambda^{\nu}} \Big\rangle,
\label{FIM_statistics}
\end{equation}
will reliably approximate the likelihood surface close to the global maximum   (here, $\mathbb{E}$ denotes an ensemble average).  The multi-variate jumps then use a product of normal distributions in each eigendirection of $\Gamma_{\m\n}$.  The standard deviation in 
each eigendirection is given by $\sigma_{\m} = 1/\sqrt{DE_{\m}}$, where $D$ is the dimensionality of the search space (in this case $D=9$), $E_{\m}$ is the corresponding eigenvalue of $\Gamma_{\m\n}$ and the factor of $1/\sqrt{D}$ 
ensures an average jump of $\sim 1 \sigma$.  This type of MCMC algorithm is commonly referred to as a Hessian MCMC and
 is known to have the highest  acceptance rate for this particular family of random walk algorithms~\cite{conf/icassp/2011}.

While the Hessian MCMC performs better than most other stochastic samplers, it can still suffer from poor mixing of the chain, especially if the parameters are highly correlated,
or the FIM is near singular.   In these cases, the acceptance rate can be prohibitively low.  One option to increase the efficiency, is to combine the MHMC algorithm with a different
type of sampler.  In previous works, it was shown that mixing the MHMC with a variant of the DE algorithm improves the acceptance rate~\cite{porter_fisher_2015,Porter:2015bea}.  

The standard DE algorithm evolves a population of solutions in parameter space via a number of simultaneous Markov chains.  In our variant of the algorithm, instead of 
multiple chains, we store a trimmed history of a single chain and use that history to propose new solutions. While not Markovian in nature, as the history grows, the chain
becomes asymptotically Markovian.  For each of the nine parameters, this then means that we have the full chain array
$x_i$ of instantaneous length $n$,  and a trimmed history array $\bar{x}_i$ with instantaneous length of $\bar{n}$, where $\bar{n} < n$.   We then construct our
proposed jump in parameter space using
\beq
x_{i+1} = x_i + \gamma \left(\bar{x}_j - \bar{x}_k \right),
\eeq
with $j,k \in U[1,\bar{n}]$,  $i\neq j\neq k$,  and $\gamma$ is the differential weight. The optimal value of this weight is given by $\gamma = 2.38 / \sqrt{2D}$ ~\cite{terbraak_2006,terbraak_2008}.   Empirically, we have found that a 2:1 ratio of MHMC to DE proposals results in an a good acceptance rate.

We implement the DEMC algorithm as follows: we run an initial $10^5$ iteration, simulated annealing based burn-in to ensure that the chain sufficiently explores the parameter space. The simulated annealing replaces the factor of $1/2$ in the likelihood expression (Eq \eqref{likelihood_definition}) with an inverse temperature $\beta = 1 / (2T)$. To set the temperature $T$ at each chain point, we use a power law  scheme ~\cite{Cornish_2006, cornish_porter_2007_1}
\begin{equation}
T = 10^{T_{0}\left(1 - i/T_{s}\right)},
\end{equation}
where $T_0 = \log_{10}(T_{ini})$ is the heat index, $T_{ini}$ is the initial temperature, $i$ is the chain iteration number, and $T_{s}=10^5$ iterations is the cooling schedule. After investigation,  to ensure our initial temperature is high enough for sufficient exploration, we set the initial temperature to $T_{ini}=50$.

During the first $5\times10^4$ iterations where we use MHMC proposals only, every accepted jump is recorded to form a history for future DE moves. Furthermore, to ensure that we are moving between modes of the solution, every $10^3$ iterations, we draw a random value from a uniform distribution $\delta\in{\mathcal U}[0,1]$. If $\delta \geq 0.5$, we propose a mode-hop of the form

\begin{eqnarray}
\iota &\rightarrow &\pi-\iota, \\
\psi & \rightarrow &\pi-\psi.
\end{eqnarray}

For the final $5\times10^4$ iterations of the burn-in, we move to the 2:1 MHMC-DE proposal mix, again recording every accepted point in parameter space for future DE proposals. The final item used in this phase is that we set the factor $\gamma=1$ every $10^2$ DE proposals. This move also encourages the chain to move between modes. Once the heat has dropped to $T=1$, we stop recording only the accepted points for the DE history, and move to a system where we record every 10th iteration of the chain, regardless of whether it is 
accepted or not.

%%%%%%%%%%%%%%%%%%%%%%%%%%%%%%%%%%%%%%

\subsection{Hamiltonian Monte Carlo}
\label{sec_description_HMC}

The HMC algorithm~\cite{duane_1987} is a stochastic algorithm that uses gradient information to efficiently sample a target distribution. As a result, it was shown that the optimal acceptance rate for a multidimensional HMC is $\sim65\%$, as opposed to $\sim26\%$ for a standard Markov Chain based sampler~\cite{Neal_2012}. 

In the HMC algorithm, the logarithm of the inverted target distribution can be interpreted as a ``gravitational" potential energy, $\mathcal{U}(q^{\mu})$, that depends on the state space variables $q^{\mu}$.  Practically,
we define the potential energy as
\begin{equation}
\mathcal{U}(q^{\mu}) = - \ln \left[ \mathcal{L}(q^{\mu}) \pi(q^{\mu}) \right],
\label{potential_energy_definition}
\end{equation}
where $\mathcal{L}(q^{\mu})$ is the log-likelihood, $\pi(q^{\mu})$ is the prior distribution, and we equate the state space variables with the parameters of the 
GW template, i.e.  $q^{\mu}=\lambda^{\mu}$.  For the set of state space variables $q^{\mu}$, we then associate a set of canonical momenta $p^{\mu}$, and define the 
 kinetic energy of the system as,
\begin{equation}
\mathcal{K}(p^{\mu}) = \frac{1}{2}M_{\mu \nu}^{-1}p^{\mu}p^{\nu},
\label{kinetic_energy}
\end{equation}
where $M_{\mu \nu}^{-1}$ is a fictitious positive-definite mass matrix.   The Hamiltonian of the system can now be written as,
\begin{eqnarray}
\mathcal{H}(q^{\mu},p^{\mu}) &=& \mathcal{U}(q^{\mu}) + \mathcal{K}(p^{\mu}), \nonumber  \\
 &=& - \ln  \left[ \mathcal{L}(q^{\mu}) \pi (q^{\mu}) \right] + \frac{1}{2}M_{\mu \nu}^{-1}p^{\mu}p^{\nu}.
 \label{Hamilton_eq}
\end{eqnarray}
\noindent As the Hamiltonian $\mathcal{H}(q^{\mu},p^{\mu}) $ defines the total energy of a system, the dynamical evolution of the system in fictitious time $t$ can then be inferred from Hamilton's equations,
\begin{eqnarray}
\frac{\dd q^{\mu}}{\dd t} &=& \frac{\partial \mathcal{H}}{\partial p^{\mu}} = \frac{\partial \mathcal{K}}{\partial p^{\mu}}, \label{theoretical_Hamil_eq_1}  \\
\frac{\dd p^{\mu}}{\dd t} &=& -\frac{\partial \mathcal{H}}{\partial q^{\mu}} = -\frac{\partial \mathcal{U}}{\partial p^{\mu}}. 
\label{theoretical_Hamil_eq_2}
\end{eqnarray}

To make the connection with Bayesian inference, we highlight the fact that the Hamiltonian has a canonical distribution over phase space, i.e.
\begin{eqnarray}
\Pi(q^{\mu},p^{\mu}) &\propto& e^{-\mathcal{H}(q^{\mu},p^{\mu})} \propto e^{-\frac{1}{2}\mathcal{U}(q^{\mu})}e^{-\mathcal{K}(p^{\mu})}\nonumber\\ 
 &\propto&\Pi(q^{\mu}) . \Pi(p^{\mu}).
 \label{Separable_proba}
\end{eqnarray}
Not only is the above density separable, but by definition $\Pi(p^{\mu})\sim\mathcal{N}(0,1)$.   This means the momentum components are independent of the state space variables 
$q^{\m}$ and each other. Thus, simply ignoring the momentum variables, the marginal distribution for $q^{\m}$ gives us a sample set which asymptotically comes from the target distribution.

In the original formulation of the HMC, the positions and momenta are evaluated discretely along a trajectory with $l$ steps, of size $\epsilon$.   A problem with this formulation is that as $\epsilon$ is
a constant in all dimensions, the algorithm does not take into account different dynamical scales in the parameters.  To compensate for
these different ranges, we use a range of step sizes $\epsilon^\m$ for the different parameters.  Using a symplectic integrator (commonly referred to as a leapfrog integrator), and assuming our mass matrix is represented by some
diagonal matrix with elements $m_\m$, we can write the expression for the scaled leapfrog equations in the following form~\cite{Neal_2012}
\begin{eqnarray}
\tilde{p}^{\mu}\left(\tau + \epsilon^{\mu}/2\right) &=& \tilde{p}^{\mu}(\tau) - \frac{\epsilon^{\mu} }{2} \frac{\partial \mathcal{U}(q^{\mu}) }{\partial q^{\mu}} \Bigr|_{q^{\mu}(\tau)}, \nonumber \\ 
q^{\mu}(\tau + \epsilon^{\mu} ) &=& q^{\mu}(\tau) + \epsilon^{\mu} \tilde{p}^{\mu}\left(\tau + \frac{\epsilon^{\mu}}{2}\right),  \\ 
\tilde{p}^{\mu}(\tau + \epsilon^{\mu} ) &=& \tilde{p}^{\mu}\left(\tau + \frac{\epsilon^{\mu}}{2}\right) - \frac{\epsilon^{\mu} }{2} \frac{\partial \mathcal{U}(q^{\mu}) }{\partial q^{\mu}} \Bigr|_{q^{\mu}(\tau + \epsilon^{\mu})}\nonumber ,
\label{eq:Scaled_leapfrog}
\end{eqnarray}
where we define the scaled momenta $\tilde{p}^{\mu} = s^{\mu}p^{\mu}$, the scaled step sizes $\epsilon^{\mu} = s^{\mu} \epsilon$, with $s^{\mu} = m^{-1/2}_{\mu}$. Without loss of generality, we now enforce that the scaled momenta are also drawn from the distribution $\Pi(\tilde{p}^{\mu}) \sim \mathcal{N}(0,1)$. As we are solving
Hamilton's equations numerically, the Hamiltonian is not strictly conserved along the trajectory.  In general, the leapfrog method introduces errors in the Hamiltonian on the order of $\mathcal{O}(\epsilon^{3})$ at each step of the trajectory and of order $\mathcal{O}(\epsilon^{2})$ for the whole trajectory.   In order for the Hamiltonian to be conserved, the values of $\epsilon^{\mu}$ should be chosen to be as small as possible to ensure conservation, but
not so small as to introduce a new computational bottleneck into the algorithm.
To make the HMC ``exact", a Metropolis evaluation is introduced at the end of each trajectory. 

The HMC algorithm then works as follows: starting at a point in parameter space $q^{\mu}_{0}$, we
\begin{enumerate}
\item Draw the scaled momenta $\tilde{p}_0^\m$ from a Gaussian distribution $\mathcal{N}(0,1)$,
\item Starting at the phase space point $(q_0^\m, \tilde{p}_0^\m)$ with Hamiltonian $\mathcal{H}(q_0^\m, \tilde{p}_0^\m)$, evolve the Hamiltonian trajectory for $l$ steps to a new phase
space coordinate $(q_0^\m, \tilde{p}_0^\m)$ with Hamiltonian $\mathcal{H}(q^\m, \tilde{p}^\m)$,
\item Accept the new phase space point with probability $\alpha = min\left(1,\exp\left[ \mathcal{H}(q^{\mu}, \tilde{p}^{\mu}) - \mathcal{H}(q_{0}^{\mu},\tilde{p}_{0}^{\mu})\right]\right)$
\end{enumerate}
As with all stochastic samplers, the HMC comes with a number of free parameters that need to be tuned for efficient sampling.  In this case: the trajectory length $l$, the
stepsize $\epsilon$ and the stepsize scaling $s^\m$ (which is dependent on our choice of mass matrix elements $m_\m$).  A further consideration, and one which has 
prevented the algorithm from wider use, is the computational cost involved in calculating the gradients of the target density in the leapfrog equations, especially if no closed-form
solution exists and the gradients have to be calculated numerically.  We will discuss
these issues, and our proposed solutions below.

%%%%%%%%%%%%%%%%%%%%%%%%%%%%%%%%%%%%%%%%%%%%%%%%%%%%%%%%%%%%%%%%%%%%%

\section{Tuning the Hamiltonian Monte Carlo}
\label{section_3}

\subsection{Introduction}
\label{section_3_1}
Before detailing the fine-tuning of the HMC algorithm, we should say a few words about our test sources, the parameterization of the HMC algorithm, our choice
of prior distributions, and how we deal with parameter space boundaries.

For our test sources, we took a sample of systems from a previous LIGO/Virgo BNS study \cite{Singer:2014qca}. To qualify for selection, we required that each source have a 
three-detector network SNR greater than a detection threshold of $\rho = 8$.  For this study, we chose ten BNS systems that reflect a range of
masses, mass ratios, inclination and sky location.  The parameters for these ten binaries can be found in Table \ref{tab:definition_sources_BNS} of Appendix \ref{BNS_table_sources}.

Choosing the correct parameter space is imperative for efficiently sampling the posterior distribution.  The gravitational waveform for non-spinning binary systems is
described by 9 parameters, i.e. $\lm=\left\{m_1, m_2, D_L, \iota, \alpha, \delta, \psi, \phi_c, t_c\right\}$, where all parameters have been described in Section II.  In general, 
$(m_1, m_2)$ are a bad choice of coordinates as the parameter space in these coordinates is degenerate.  Likewise, using the symmetric mass ratio $\eta$ is also
a bad choice due to a physical cutoff at $\eta=1/4$ (i.e. $m_1=m_2$).  In past studies~\cite{porter_2009,Cornish_2006,porter_fisher_2015,Cornish_2005_data}, we found that using $(\mathcal{M}_c, \mu)$, where $\mu = M\eta$ is the 
reduced mass, produces well mixing chains.  To reduce the dynamic range in parameter space, and to lower the condition number for the FIM, we also found that using
$(\ln\mathcal{M}_c, \ln\mu, \ldl, \ln t_c)$ work better that using the bare coordinates.  Finally, we also observed that our chains mix better using the latitude and longitude
$(\theta, \phi)$ of the source, rather than right ascension and declination.  Pulling this together, we define our parameter space coordinates
\begin{equation}
q^{\mu} = \{\cos \iota, \phi_c, \psi, \ldl, \ln \mathcal{M}_{c},\ln\mu, \sin \theta, \phi, \ln t_c\}.
\end{equation}

For our prior distributions, we chose uninformative (flat) priors. We chose our priors in $(\ln \mathcal{M}_{c},\ln\mu)$ such that they correspond to individual masses in the range $m_i\in[1,2.3]\,M_{\odot}$. Our distance prior is flat in $\ldl$ corresponding to $\ldl\in[10^{-6}, 200]$ Mpc, where the lower bound corresponds to the distance to 
the M31 galaxy, and the upper bound is defined by the design sensitivity BNS range of aLIGO. For the chirp-time, our prior is flat in $\ln t_c$, such that  $t_c\in[\overline{t}_{c} - 5, \overline{t}_{c} + 5]$ seconds where $\overline{t}_{c}$ is the true value of coalescence time. 
Finally, for the priors in our angular parameters, we choose  $(\cos  \iota,\sin \theta)\in[-1,1]$, $(\phi,\varphi_c)\in[0,2\pi]$ and $\psi\in[0,\pi]$.

The last point of discussion is how each algorithm deals with boundaries in parameter space.  In the case of the DEMC algorithm we can use 
simple rejection sampling to ensure that the proposed points are inside both the astrophysical (e.g. $\eta\leq1/4$) and prior boundaries.  We
could also use rejection sampling for the HMC algorithm, but it can be very costly.  As each trajectory takes a non-negligible amount of time to
generate, rejecting a trajectory that wanders outside a boundary is a waste of computing cycles.  To circumvent this problem, we impose a reflective boundary around our 
parameter space for those parameters for which a simple re-mapping does not exist.  For the luminosity distance and the mass parameters, 
this reflective boundary condition reverses the direction of the trajectory by negating the sign of the associated canonical momentum.   
For the remaining parameters, we ensure that at each step of a trajectory, their values are within their natural range and remap them if necessary.

\subsection{Tuning the Mass matrix}
\label{mass_matrix_section}
In Ref. \cite{Porter_2013}, a proposal for the mass matrix was made to reflect the different dynamical ranges for each parameter.  In this case, and under the assumption that the
FIM is a sufficiently good quadratic approximation to the local curvature of the log-likelihood,  the diagonal elements of the mass matrix
could be set by using the variance predicted by the inverse of the FIM, i.e. $C_{\mu\nu} = \Gamma^{-1}_{\mu\nu}$, giving
\begin{equation}
m^{-1}_{\mu} = C_{\mu\mu} = \left(\sigma^{FIM}_{\mu} \right)^{2},
\label{mass_matrix_components}
\end{equation}
where $\sigma^{FIM}_{\mu} $ are the standard deviations predicted by the FIM, which allows us to define a scaling factor $s^{\mu} = \sigma^{FIM}_{\mu}$.

This choice of mass matrix provided good results during our initial tests on a single test case, with both good exploration and acceptance rate. However, when we applied the algorithm to other test sources, we found that the acceptance rates were unnaturally low. Upon investigation, we found that in these cases, the FIM was close to being singular and predicted scales in certain eigendirections that were larger than their natural scales, e.g. $s^{\ldl} = \Delta D_L/D_L > 1$. For other parameters, such as $(\psi, \phi_c)$, this had a negligible impact on the mixing of the chain as we could easily re-map the solution back into the
natural range.  However, when this happened to parameters such as $(\ldl, \cos\iota)$ it caused the Hamiltonian along the trajectory to diverge quite rapidly, leading to all
 trajectories being rejected. As a solution, we constrained the scales in the problem eigendirections by setting
\begin{eqnarray}
s^{\cos \iota} &=& 1\,\,\,\,\,       \text{if} \,\,\,\,\,s^{\cos \iota} \geq 1, \\
s^{\phi_{c}} &=& \pi ,\,\,   \,\,\,\,     \text{if} \,\,\,\,\,s^{\phi_{c}} \geq \pi, \\
s^{\psi} &=& \dfrac{\pi}{2} ,\,\,   \,\,\,     \text{if} \,\,\,\,\,s^{\psi} \geq \dfrac{\pi}{2}, \\
s^{\ldl} &=& 0.5 \,\,\,\,   \,    \text{if} \,\,\,\,\,s^{\ldl} \geq 0.5,
\end{eqnarray}
corresponding to a $50\%$ error-prediction for these four parameters. 

%%%%%%%%%%%%%%%%%%%%%%%%%%%%%%%%%%%%%%%%%%%%%%%%%%%%%

\subsection{Tuning the step size}

While a fixed value of $\epsilon = 2.5 \times 10^{-3}$ produced HMC chains with good mixing, this option is not necessarily the optimal choice. Other more generic HMC variants attempt to tune the free parameters on the fly \cite{NUTS}. However, as we are trying to develop an algorithm specifically for GW astronomy, we decided to take the time to empirically optimise the step size.  Initially, we tried a number of different fixed values of $\epsilon$, but were unable to clearly determine one specific value that satisfied the
necessary criteria for a well mixing chain.

Following Ref~ \cite{Neal_2012}, we finally decided to draw $\epsilon$ from a normal distribution with a mean of $5\times10^{-3}$, and a standard deviation of $1.5 \times 10^{-3}$, ensuring that $\epsilon\in[10^{-3},10^{-2}]$.  The idea here is that trajectories with small values of $\epsilon$ will always be accepted,
as their Hamiltonians will always be conserved.  However, the end point of the trajectory may not be very far from the starting point in parameter space, so exploration is local.  On
the other hand, trajectories with large $\epsilon$ have a higher probability of being rejected. But in those cases where they are accepted, the exploration is wider and the trajectory end points are
more likely to be statistically independent of the starting point.  In all test cases, we found that drawing the stepsize from a normal distribution produced more statistically independent chains than
those with a fixed value.

%%%%%%%%%%%%%%%%%%%%%%%%%%%%%%%%%%%%%%%%%%%%%%%%%%%%%

\subsection{Tuning the trajectory length}
If the value of $l$ is properly tuned, the HMC should produce arcs in phase space.  However, for a given value of $\epsilon$, if the value of the trajectory length $l$ is too small, the arc will
be such that the start and end points of the trajectory will be so close together that the chain random walks through parameter space. If the value of $l$ is too large, the trajectory generates a circle in phase space, and once again the end point is sufficiently close to the starting point that the chain is again a random walk . To avoid the situations described above and ensure a well mixing chain, and also taking into account the variable size of $\epsilon$, we draw the trajectory length from a uniform distribution (that we will define later in this article).

%%%%%%%%%%%%%%%%%%%%%%%%%%%%%%%%%%%%%%%%%%%%%%%%%%%%%

\subsection{Calculating the gradients of the log-likelihood}
When evolving the Hamiltonian trajectory via the leapfrog equations, the most computationally expensive task is evaluating the gradients
of the log-likelihood.  Even with the simple TaylorF2 waveform, there is no closed-form solution for the gradients, and hence, they need to be calculated
numerically.  In general, using central differencing, 17 waveform generations
are needed to calculate the 9 log-likelihood gradients at each step of the trajectory (not 18 as $\partial \tilde{h}/\partial \ldl = -\tilde{h}(f)$).

While we can use a mixture of analytical and numerical differencing techniques,
we still require 9 waveform generations at each trajectory step.  This is a major computational bottleneck for the algorithm.  As an example, the average
time to generate a 200-step trajectory on a 2.9 GHz Intel i5 processor was close to 6 seconds.  This corresponds to a runtime of close to 10 weeks for a full 
$10^{6}$ trajectory HMC run.  In some cases, we may want to use more complicated, and hence expensive waveforms, meaning that the runtime could increase
by an order of magnitude or more.  These timescales clearly prohibit the use of the HMC as a viable sampler.  To get around this problem, we need to 
generate the log-likelihood gradients faster.

The working solution proposed in \cite{Porter_2013} divides the algorithm into three parts, which for future reference, we will define as Phases I-III.  In Phase I, 
the algorithm begins by running a small number of 
expensive numerical trajectories $(\sim10^3)$.  At each point along the trajectory, the coordinate positions and values of the derivatives of the log-likelihood are recorded.  If a trajectory is accepted, 
then the individual coordinate positions are kept for future use, otherwise they are rejected.  In Phase II, the recorded positions are used
to fit the following analytic function in each of the nine dimensions:
\begin{equation}
f(q^{\mu}) = \sum_{i=1}^{D} a_{i} q^{i} + \sum_{j=1}^{D} \sum_{k=j}^{D} a_{jk} q^{j} q^{k} + \sum_{l=1}^{D} \sum_{v=l}^{D} \sum_{w=v}^{D} a_{lvw} q^{l} q^{v} q^{w}.
\label{eq:gradient_approximation}
\end{equation}
Here, $D$ is the dimension of the parameter space,   and $a_{i}$, $a_{jk}$ and $a_{lvw}$ are the fit coefficients.   Phase I normally takes a few hours to run.

In Phase II of the algorithm, the fit coefficients are computed using a least squares method.   As well as the nine $a_i$ coefficients,
the  $a_{jk}$ and $a_{lvw}$ represent the independent coefficients of a $D \times D$ symmetric matrix, and a $D\times D\times D$ symmetric tensor respectively. In the non-spinning case where $D = 9$,
 a total number of $220$ coefficients are needed  for each gradient.  In general, during the first Phase I,
we construct an $n\times 9$ matrix of accepted coordinates, where $n\sim10^5$.  The least squares method requires an inversion of this matrix.   
Originally, we used a singular value decomposition (SVD) to invert the matrix.  As ``tall and skinny"
matrices are notoriously ill-conditioned for inversion, it takes over an hour to fit all nine gradient approximations using a SVD.    At a later point in the study, we substituted the
SVD method with a QR decomposition.  This reduced the fitting time to less than ten minutes.

In Phase III, the numerical trajectories from Phase I are replaced with trajectories generated using the analytic fit.  It is important to highlight, that if
the gradient approximation is good, we can essentially define a ``shadow" potential which is almost identical to the true potential.  The algorithm then proceeds as follows: the starting point
for each trajectory (where the Hamiltonian has been calculated via a waveform generation) lies on the true potential.  To evolve a trajectory, we then move off the true potential and onto the shadow potential, using the analytic
gradient approximation for $l$ steps.  At the end of the trajectory, we move back onto
the true potential, generate a waveform, calculate the Hamiltonian, and compare the Hamiltonians at the start and end points of the trajectory via the Metropolis-Hastings ratio.  In general, this
results in an acceleration of factors close to $10^3$ with no loss in accuracy.

\begin{figure}
\epsfig{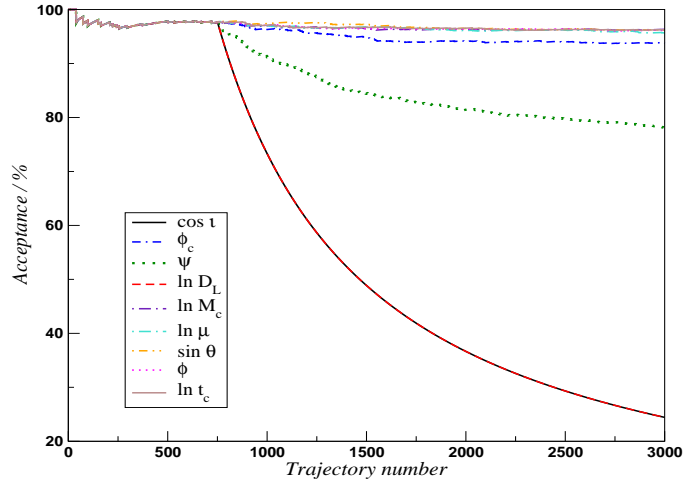}
%\caption{Acceptance rate depending on the number of trajectories for a set of simulations where we ran $750$ initial numerical trajectories and then used fitted gradient with respect to the parameter denoted in the legend while keeping the other eight gradients computed numerically. For $\cos \iota$ and $\ln D_{L}$, the curves lie on top of each other and none of the trajectories are accepted during phase II.}
\caption{Acceptance rates of single-parameter analytic gradient trajectories.  While the acceptance rates for six of the parameters stays almost constant
between Phase I and III, there is a rapid decrease in acceptance rate for the parameters $\left\{\cos \iota, \ln D_{L}, \psi\right\}$.  In the plot, the curves for $\cos \iota$ and $\ln D_{L}$
lie on top of each other.}
\label{8num_1analytical}
\end{figure}

\begin{figure}[th]
\epsfig{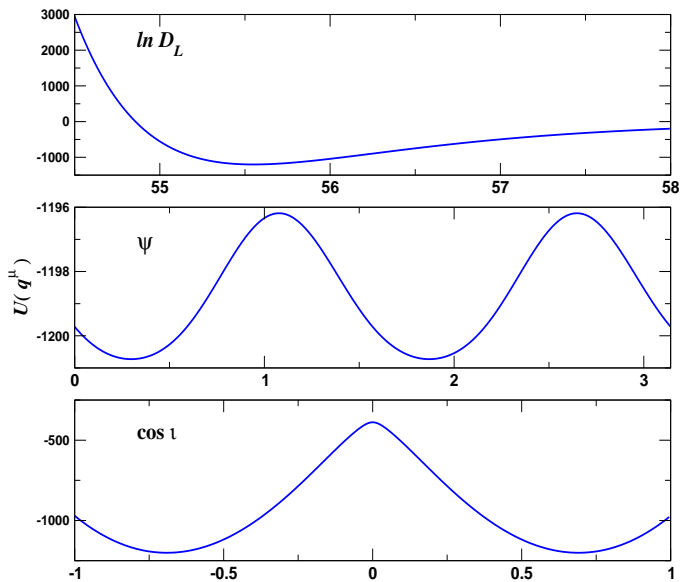}
\caption{Potential wells, $\mathcal{U}(q^\m) =  - \ln  \left[ \mathcal{L}(q^{\mu}) \pi (q^{\mu}) \right]$,  for the three problem coordinates $\{\ln D_{L}, \psi, \cos \iota  \}$, for BNS1.  On investigation, we can
see that the barriers between modes in $\psi$ are very shallow.  However, both $\ldl$ and $\cos\iota$ have high energy barriers which can be a problem for the HMC algorithm.}
\label{fig:potentials}
\end{figure}

%As our prior distributions are uniform, the derivatives of the posterior distribution in Eqns \eqref{Scaled_leapfrog} become dominated by the gradients of the log-likelihood expressed as,
%\begin{equation}
%\dfrac{\partial \ln \mathcal{L}}{\partial q^{\mu}} =  \Big\langle s \Big\rvert  \dfrac{\partial h}{\partial q^{\mu}} \Big\rangle - \Big\langle h \Big\rvert \dfrac{\partial h}{\partial q^{\mu}}  \Big\rangle.
%\end{equation}
%If we evaluate these gradients with numerical differencing, at each step of the trajectory we would need to generate 17 waveforms since the gradient of the log-likelihood with respect to $\ln D_{L}$ is the negative of the waveform itself. However, given the simple form of the TaylorF2 waveforms, we managed to derive analytical expressions for most of the derivatives with the exception of the derivatives of the beam pattern functions that were computed using a forward-differencing scheme,
%\begin{equation}
%\dfrac{\partial F^{+,\times}}{\partial q^{\mu}} \approx  \dfrac{ F^{+,\times}(q^{\mu} + \delta q^{\mu}) - F^{+,\times}(q^{\mu})  }{ \delta q^{\mu}},
%\end{equation}
%where we set $\delta q^{\mu} = 10^{-6}$. With this method, we reduced the total number of waveform generation at each step of a trajectory down to 9. However, the computation time to compute the gradients is still a major bottleneck for the algorithm. As an example, we found that the average computation time to generate a 200-step trajectory on a 2.9 GHz Intel i5 processor was close to 6 s, which corresponds to a computation time close to 10 weeks for a full $10^{6}$ trajectories run.  

\begin{figure}
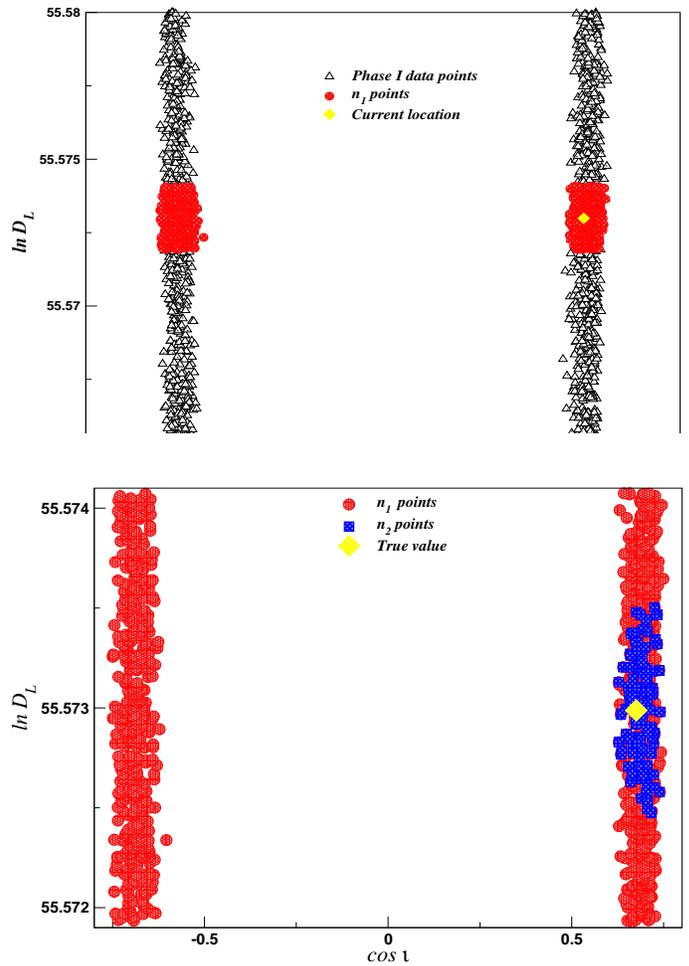

\begin{center}
%\begin{minipage}[b]{.48\linewidth}
\centering
 \epsfig{file=Illustration_n2_New.eps, width=0.5\textwidth, height=2.5in}
%\end{minipage} \hfill
%\begin{minipage}[b]{.48\linewidth}
\centering
  \epsfig{file=Dispersion_LogD_Article_New_n2.eps, width=0.5\textwidth, height=2.5in}
%\end{minipage}
% \caption{(Left) Illustration of the problem arising with the selection method of the $n_{1}$ points for the fit of the gradient of $\ln D_{L}$ using look-up table. We plot in the $\{\cos(\iota), \ln D_{L}\}$ two-dimensional surface the full set of fit points (blue), the $n_{1}$ points for the fit (red) and the true solution (green). (Right) Illustration of the selection of the $n_{2}$ points to fit the gradient of $\ln D_{L}$ at the position $q^{\mu}_{fit}$ represented in orange. The points are plotted in the $\{ \cos \iota, \ln D_{L} \}$ two-dimensional surface.}
\caption{Methodology for approximating the log-likelihood gradients with look-up tables, using $\partial \ln 
\mathcal{L}/ \partial \ln \bar{D}_{L}$ at $\bar{q}^{\m}$ (yellow diamond) as an example.  (Top) Using the pool of recorded coordinates from Phase I
(blue triangles), identify $n_1$ closest points in the $D_L$ coordinate (red circles) symmetric around the point of interest (green square). (Bottom) From the $n_1$ points, identify the subset of
points $n_2$ that are not only closest to the point of interest, but that are also located on the same mode (orange squares).  These $n_2$ data points are then used to construct
a linear approximation of the gradient.}
 \label{fig:illustration_selection_n1_n2}
\end{center}
\end{figure}

With minor modifications, we implemented the HMC algorithm defined above for BNS parameter estimation, and began tests on a single test source.  In Phase I, we used 
750 initial numerical gradient trajectories, with $l=200$.  The acceptance rate during this phase was close to $96\%$ , generating $\sim 1.5 \times 10^{5}$ coordinate points
in parameter space.   We then used these points to fit the coefficients in the gradient approximation, to be used in Phase III.  Upon entering Phase III, we noticed almost
immediately that the HMC was getting stuck and the acceptance rate fell to zero quite rapidly.  Starting the algorithm from different points in parameter space resulted
in the same behaviour.

To understand why the algorithm was not performing as expected, we individually tested the approximate gradients for each of the  nine parameters, i.e.  where one parameter used approximate gradient trajectories, while the other eight used numerical gradient trajectories.  In Fig~\ref{8num_1analytical}, 
we plot the acceptance rates for each of the individual parameter simulations as a function of trajectory number.   For six of the parameters, $\{ \ln \mathcal{M}_{c}, \ln \mu ,\phi_{c}, \ln t_{c}, \sin (\theta),\phi \}$, we see that the acceptance rate stays almost constant between Phase I and III, with a value close to $96\%$.  In these cases, the gradient approximation is working well and ensures that the conservation of the Hamiltonian along the trajectory is almost as good as when using numerical gradients.
For the final three parameters, $\{\ldl, \psi, \ci  \}$, we see that the acceptance rates drop off almost immediately as the chains hardly move in Phase III.  We note that due to their high correlation, the acceptance
rates for $D_L$ and $\ci$ drop off at exactly the same rate.

%To understand what is going on, it is worth investigating the potential energy wells for these three parameters (see Fig.~\ref{fig:potentials}).  Going from
%top to bottom, $\ldl$ has a single potential well with a high energy barrier on one side.  On the other hand, for $\psi$ we have a number of mini-wells across 
%the parameter range.  However, on inspection, the potential barriers are shallow enough that moving from one well to another is not really an issue.  The main problem for the HMC algorithm comes
%from the potential well for $\cos\iota$.  We see two clear potential wells, with a large energy barrier in between.  This constitutes
%a major problem for the HMC algorithm in general as it is known to have difficulties traversing high energy barriers.  Upon inspection, we observed that during Phase I, most of the transitions between
%the two potential wells was via a remapping of the parameter as it exited the boundaries at -1 and 1.  This left a large part of the parameter space at $\cos\iota\sim0$ which was unsampled during Phase I,
%causing the approximate gradient to fail in Phase III when trajectories entered this region.  In an attempt to find a solution for the problem gradients, we tried higher-order polynomial fits (quartic, quintic), splitting the pool of points in two sets 
%depending on the value of $\cos \iota$ and using radial basis functions \cite{skala_2016}. None of these methods provided an acceptable solution.   

To answer what is going on, it is worth investigating the topology of the potential wells for these three parameters.  In Fig.~\ref{fig:potentials}, we plot the
potential wells at a slice in parameter space, where we keep eight of the parameters at the true values, and allow the ninth to vary.  Going from top to 
bottom, we see that $\ldl$ has a single potential well with a high energy barrier on one side.  On the other hand, both $\psi$ and $\ci$ have two separated
potential wells.  If we look at the value of ${\mathcal U}(\psi)$, we see that the wells are quite shallow, with a depth of $\Delta{\mathcal U}(\psi)\approx5$.
An energy barrier this shallow does not constitute a problem for the HMC, and the algorithm passes quite easily between the two potentials.  On the 
other hand, both of the potentials in $\ci$ are separated by a much higher energy barrier with  $\Delta{\mathcal U}(\ci)\approx750$.  This is a problem
for the algorithm.

Upon inspection, we found that the majority of the Phase I trajectories moved between the two $\ci$ potentials via a remapping of the parameter upon
exiting the boundaries at $\pm1$.  As the height of the potential is a function of parameter space position, the other transitions took place across the
energy barrier at a point where it was shallow enough for the algorithm to cope with.  The result was that this left a large part of the parameter space around
 $\ci\sim0$ which was unsampled during Phase I,  causing the approximate gradient to fail in Phase III when trajectories entered this region.  In an attempt 
 to find a solution for the problem gradients, we tried higher-order polynomial fits (quartic, quintic), splitting the pool of points in two sets 
depending on the value of $\cos \iota$ and using radial basis functions \cite{skala_2016}. None of these methods provided an acceptable solution. 

The solution we finally settled for was to use a local linear fit based on ordered look-up tables. To illustrate this method, we take the example of fitting $\partial \ln 
\mathcal{L}/ \partial \ln D_{L}$ at position $\overline{q}^{\mu}$.  If we proceed naively,
given the $N$ recorded data points from Phase I, we order each look-up table for $\left\{\ci, \ldl, \psi\right\}$.  Choosing the $\pm M/2$ data points
in the table closest to $\bar{D}_L$ (such that
$M << N$) , we use a linear fit to approximate the gradient $\partial \ln \mathcal{L}/ \partial \ln \bar{D}_{L}$.  However,  the top cell of
Fig.~\ref{fig:illustration_selection_n1_n2} demonstrates why this approach does not work. If we plot a slice in $\left\{\cos\iota, \ldl\right\}$ space, where the current point is
represented by the yellow diamond, we can see that due to the bi-modal distribution in $\ci$, the $M$ closest points to $\bar{D}_L$ in the look-up table
are spread across 
both modes.  As the linear fit requires locality of the points, the local gradient approximation then fails.

To solve this problem, we use a two-step selection process, as illustrated in Fig.~\ref{fig:illustration_selection_n1_n2}.  In the top cell, starting with the pool of $N$ accepted 
data points from Phase I (black triangles), we identify the closest tabular value in luminosity distance, $D^*_L$, to the point of interest, $\bar{D}_L$.  We then take the $\pm M/2$ points 
in the table around $D^*_L$ to construct our initial set of $n_1=M+1$ selection points (red circles).  At this point, we have now identified the the closest $n_1$ values of
luminosity distance to  $\bar{D}_L$.  However, the values of the other eight parameters are still most likely to be spread across multiple modes.

\begin{figure*}[t]
\epsfig{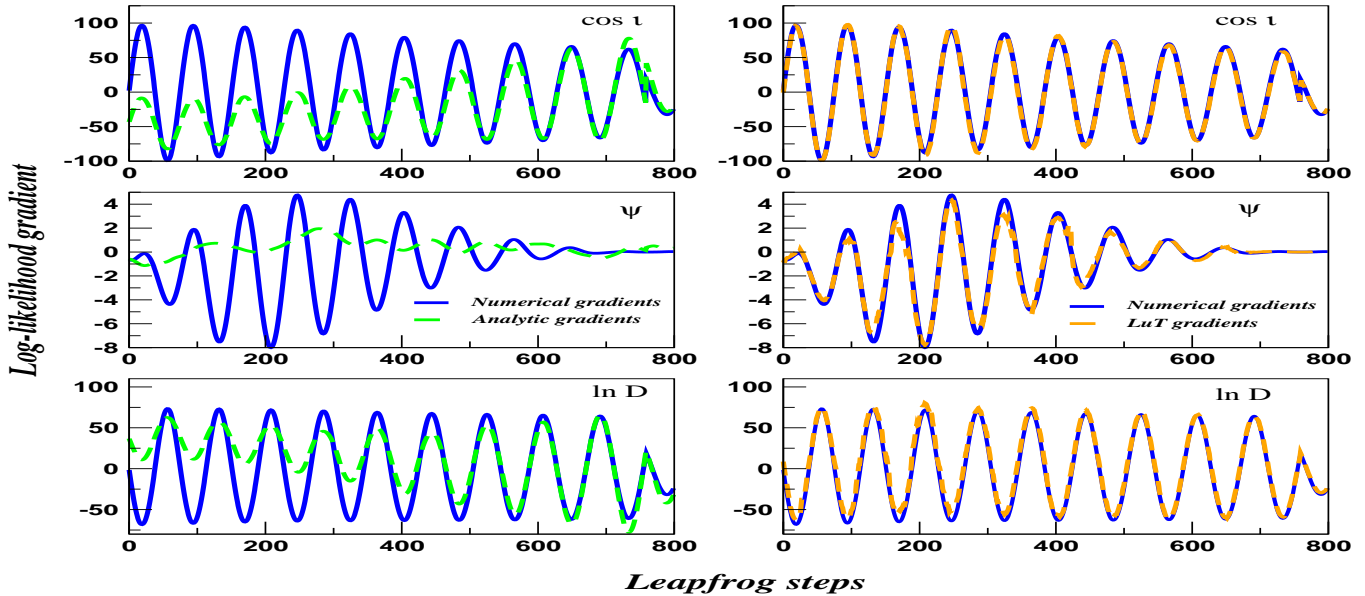}
\caption{Comparison of the numerical and approximate gradients of the log-likelihood with respect to $\{\cos \iota, \psi, \ln D_{L} \}$ for a single trajectory with $l=800$. The left panels 
compare the numerical and analytical gradient approximation.  We see a clear mismatch between the two.  The panels on the right compare the
numerical and look-up table gradients.  In this case, the approximate gradients almost perfectly match the numerical values.}
\label{fig:num_vs_approx}
\end{figure*}

As represented in the bottom cell of Fig.~\ref{fig:illustration_selection_n1_n2}, the next step is to identify
a sub-set of points, $n_2$ (such that $n_2 < n_1$),  that are not only on the same mode, but are truly local to the point of interest (blue squares).   
To ensure this criterion is met, rather than having to define a metric in the full $9D$ space, it turns out that only the metric in the $3D$ sub-space of problem parameters
is important.  So continuing our example, to confine the tabulated data points to a single mode, local around $\bar{D}_L$, we use a scaled Euclidean distance of the
form
\beq
%\parallel q^{\mu}_{i} - \bar{q}^{\mu} \parallel_{\ln\bar{D}_L}^{2} = \left( \frac{q^{\cos\iota}_{i} - \bar{q}^{\cos\iota} }{s^{\cos\iota}} \right)^{2} + \left( \frac{q^{\psi}_{i} - \bar{q}^{\psi} }{s^{\psi}} \right)^{2},
\parallel q^{\mu}_{k} - \bar{q}^{\mu} \parallel^2 = \left( \frac{(\cos\iota)_{k} - \cos\bar{\iota}}{s^{\cos\iota}} \right)^{2} + \left( \frac{\psi_k - \bar{\psi} }{s^{\psi}} \right)^{2},
\eeq 
where $q_k^\m$ are the $n_1$ points in the initial selection, and $s^{\mu}$ are the scalings defined in Sec. \ref{mass_matrix_section}. Similar expressions are used when calculating the 
gradients with respect to $\cos\iota$ and $\psi$ with the appropriate permutations.  Using the $n_2$ data points, we use a linear fit to approximate the gradient.
Empirically, we found optimal values of $(n_{1},n_{2}) = (2000,100)$ data points.

In Fig.~\ref{fig:num_vs_approx} we demonstrate how well this new gradient approximation works.  On the LHS of the plot, we compare the values of the numerical
gradients (blue) against the analytic gradient approximation given by Eqn.~(\ref{eq:gradient_approximation}) (green) along a trajectory, for the three problem coordinates $\left\{\ci, \ldl, \psi\right\}$.  We clearly see that the analytic 
approximation has trouble reproducing the true gradient values.    On the RHS, we now plot the values of the numerical gradients against the new approximation using ordered look-up tables (orange).
In this case, we have an almost perfect reproduction of the numerical values.  While not as fast as the analytic approximation, calculating the gradients based on look-up tables greatly improves the efficiency of the algorithm.   In general, we
we were able to reduce the average computation time for a single $200$ step trajectory from $6$ seconds to $66$ ms.  The new approximation also kept
 the acceptance rate almost constant between Phase I and III.

We should highlight that in order for the new gradient approximation to work well, we need to ensure that the look-up tables contain a sufficient density of data points.  There
are two practical ways of doing this:  the first, and most computationally expensive way, is to simply increase the number of numerical gradient trajectories in Phase I.  However,
as the number of required data points is going to be dependent on the complexity of the posterior distribution, and hence different sources, we have no way of knowing a-priori
just how many data points we need.  A second, and potentially cheaper approach, is to populate the look-up tables on-the-fly.  In the final version of the algorithm, we use a mixture
of both methods.

To ensure a smooth transition from Phase I to Phase III, we increased the number of numerical gradient trajectories in Phase I from 750 to 1500.  This phase now takes between
2.5-5 hours and generates $\sim3\times10^5$ data points for fitting the analytic gradients.  As we pointed out above, we are aware that this will not be sufficient to fully describe the
posterior gradients, especially for those parameter spaces with a complex geometry.  However, any further increase in the initial number of numerical gradient trajectories
has a large impact on the overall runtime of the algorithm.  To ensure accuracy and efficiency, the final algorithm structure is as follows:

\begin{itemize}
\item Phase I: Run 1500 trajectories using numerical gradients, with $l=200$.  Record the visited coordinates of each accepted trajectory.
\item Phase II: Use a QR decomposition to fit the 220 coefficients in $f(q^\m)$ for each of the parameters $\left\{\ln\mathcal{M}_{c}, \ln \mu ,\phi_{c}, \ln t_{c}, \sin \theta,\phi\right\}$.  Build 
the ordered look-up tables for $\left\{ \cos \iota, \psi, \ln D_{L} \right\}$.
\item Phase III: A well tuned HMC algorithm has an acceptance rate (AR) of $\sim65\%$.  To ensure that the algorithm mixes well, Phase III is composed of a number of features.
\begin{enumerate}
\item If $AR\geq65\%$, each trajectory uses analytic/look-up approximation gradients with $l\in U[50,100]$.
\item If $50\% \leq AR < 65\%$, we use hybrid trajectories combining analytical gradients for $\left\{\ln\mathcal{M}_{c}, \ln \mu ,\phi_{c}, \ln t_{c}, \sin \theta,\phi\right\}$ and numerical
gradients for $\left\{ \cos \iota, \psi, \ln D_{L} \right\}$, with  $l\in U[50,100]$.  These trajectories are necessary for areas in parameter space where the look-up tables do not contain a sufficient density
of data points to properly approximate the gradients with respect to $\left\{ \cos \iota, \psi, \ln D_{L} \right\}$.  
\item if $AR < 50\%$, we revert to using full numerical gradient trajectories with $l\in U[50,100]$.  In general, this step is needed if we suddenly visit a part of the parameter space for the first time.  
\item To ensure that chain does not get stuck, if three consecutive trajectories are rejected, we alternate between hybrid/numerical trajectories with  with $l\in U[20,100]$ until the chain
unsticks itself
\item After each hybrid/numerical trajectory, we update and re-sort the look-up tables
\item Every $10^5$ trajectories, with the new accumulated data points, use the QR decomposition to again fit the 220 coefficients in $f(q^\m)$ for the six well behaved coordinates.
\end{enumerate}
\end{itemize}

\section{Results and discussion}
\label{section_4}

In this section we present the results of an application of the HMC algorithm to a set of ten BNS test-sources.  For the sake of comparison, we compare our results with the
DEMC algorithm described earlier.  For each test-source, the DEMC/HMC algorithms were run for $10^6$ iterations/trajectories respectively.  While we know that this is not long enough for the DEMC algorithm to fully converge to the target distribution, it does allow
us to make an apples-to-apples comparison between the two algorithms.

%\subsection{Convergence Diagnostics}

\begin{figure}[th]
\epsfig{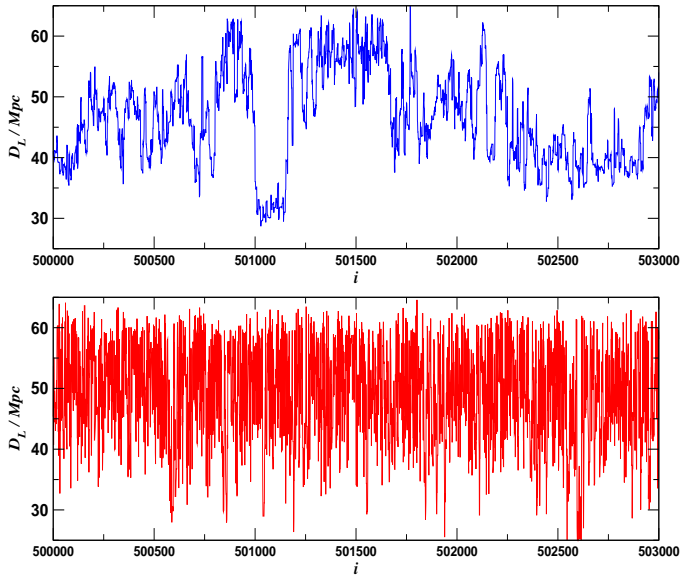}
\caption{A snapshot of the DEMC (top) and HMC (bottom) chains for $D_L$ using BNS1.  We can see that the HMC chain mixes faster, indicating much
lower serial correlation between successive draws.  This allows a much more efficient exploration of the parameter space.}
\label{fig:mixing}
\end{figure}

\begin{figure*}
\begin{center}
\begin{minipage}[b]{.48\linewidth}
\centering
\epsfig{file=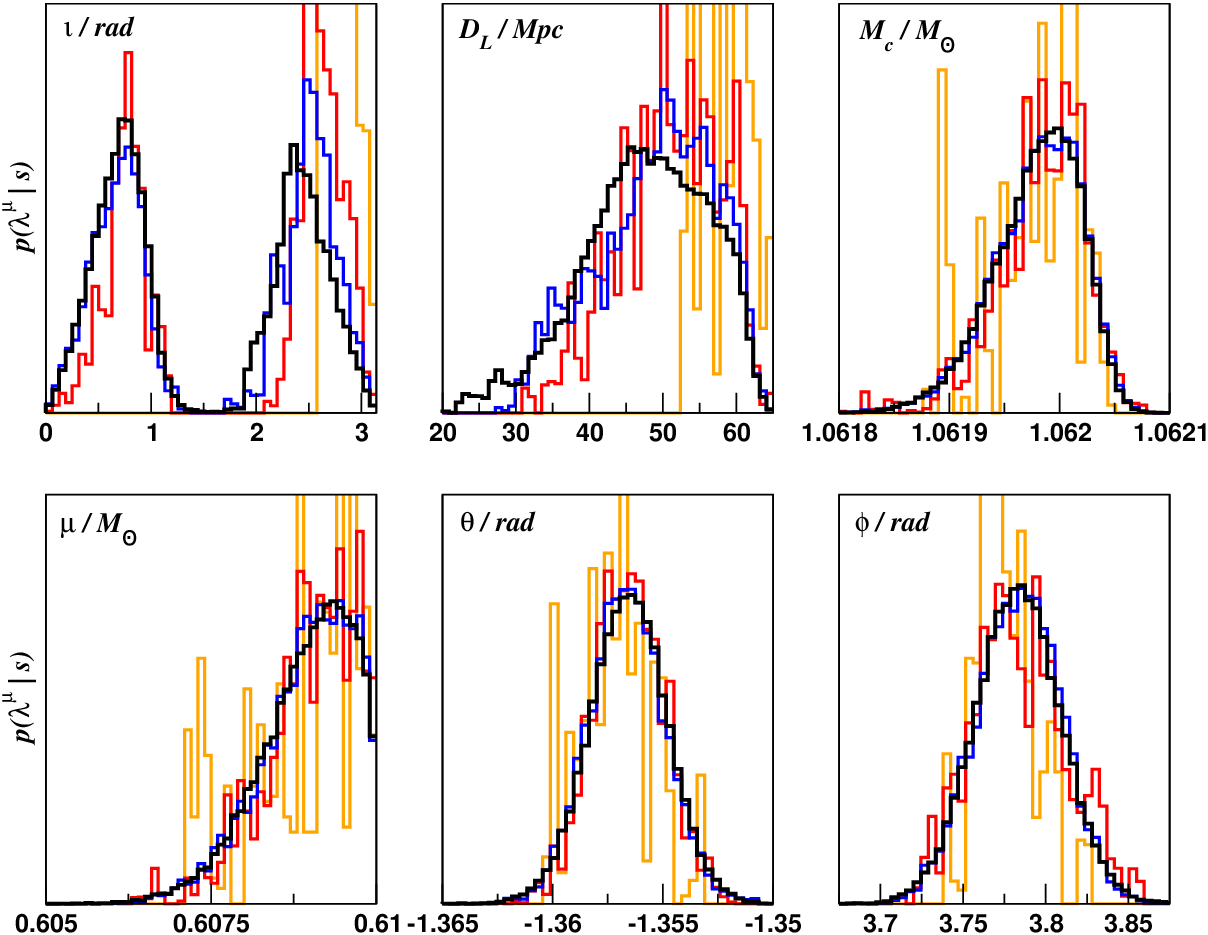, width=1.0\textwidth, height=2.5in}
\end{minipage} \hfill
\begin{minipage}[b]{.48\linewidth}
\centering
\epsfig{file=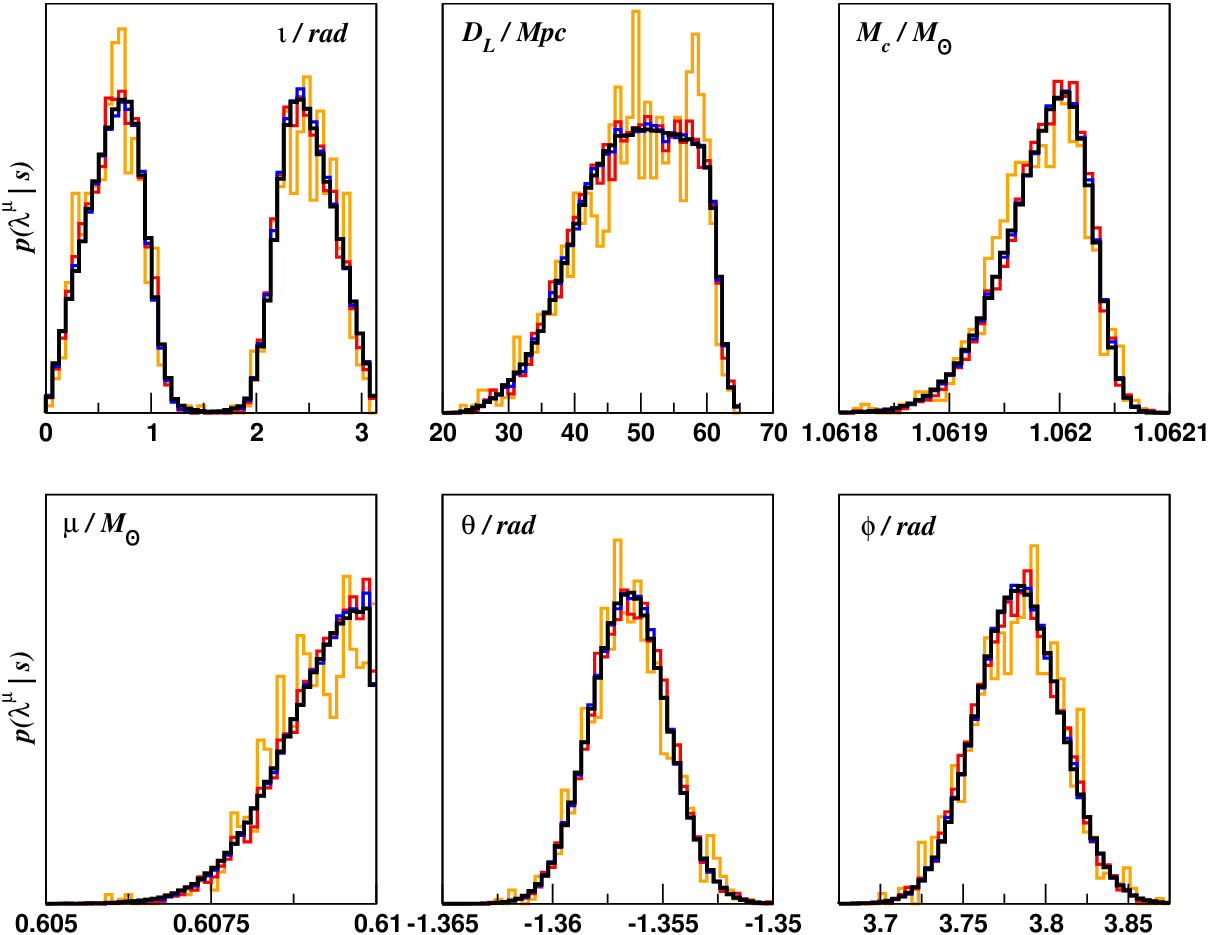, width=1.0\textwidth, height=2.5in}
\end{minipage}
 \caption{Evolution of the marginalised posterior distribution for BNS1, using a DEMC (left) and HMC (right) algorithm at snapshots of $10^3$ (orange), $10^4$ (red), $10^5$ (blue) and $10^6$ (black) iterations/trajectories, for the parameters $\{\iota, D_L, M_c, \mu, \theta, \phi\}$.}
\label{fig:bns1_conv_post}
\end{center}
\end{figure*}

There is no single universal criterion to assess the convergence of a Markov chain. While some numerical methods for convergence estimation exist, they can sometimes require running multiple chains~\cite{Geweke92evaluatingthe,Raftery92howmany,gelman1992}, adding extra cost to the analysis.  Furthermore,
as no one test can definitively say that a Markov chain has converged, we use a number of different options when diagnosing the performance of the algorithms.

The simplest test is a visual inspection of the chains themselves.  In Fig~\ref{fig:mixing}, we plot a snapshot of the $D_L$-chain, for BNS1, using
the DEMC (top) and HMC (bottom) algorithms.  For the DEMC, we observe a slow mixing of the chain, leading to a random-walk behaviour in parameter
space.  The evolution of the chain suggests high serial correlation between successive draws, thus leading to very slow exploration.  In contrast, the HMC
chain displays a good level of mixing, with very rapid exploration in parameter space.  This behaviour was common for all parameters and for all sources in our study.

Our next test is an inspection of  the convergence of the marginalised posterior distribution as a function of chain length.  Again, using BNS1 as an example, 
in Fig.~\ref{fig:bns1_conv_post}, we plot snapshots of the posterior convergence for the parameters $\{\iota, D_L, \mathcal{M}_{c}, \mu, \theta, \phi\}$ after $10^3$ (orange), $10^4$ (red), $10^5$ (blue) and $10^6$ (black) iterations/trajectories, using the DEMC (left panel) and HMC (right panel) algorithms.   We should note here that for clarity, the true values are not plotted against the distributions.  However, for reference, the true values are
 $\{\iota, D_L, \mathcal{M}_{c}, \mu, \theta, \phi\} = \{0.8029, 43,  1.062, 0.60996, -1.356, 3.777\}$, where luminosity distance is in Mpc, the masses are in solar masses, and the inclination and sky angles are in radians.

%If we first focus on the DEMC chain, as expected, with only $10^3$ samples, the distribution is highly peaked, and in the case of inclination, is concentrated on one mode of the solution. It is only from $10^5$ iterations onwards that we observe the first real signs that we are beginning to converge to the target distribution for $\{\mathcal{M}_{c}, \mu, \theta, \phi\}$ as when compared to the posterior using $10^6$ iterations. For $\{\iota, D_L\}$, we see that our distribution is still very peaked, and also still quite shifted from the distribution obtained with the $10^6$ samples. In the case of the HMC chain, the posterior distribution for $10^{3}$ trajectories already contain the general features of the posterior distributions such as the bimodality in inclination. At $10^{5}$ trajectories, we find that the HMC chain is already displaying strong signs of convergence with no visible differences with the marginalised posterior distributions obtained with $10^{6}$ trajectories.

First, focusing on inclination, we see that the DEMC algorithm requires between $10^3$ and $10^4$ iterations to begin sampling both modes.  As a display of the slow convergence of the chain, we also see a large 
difference between the posterior distributions at $10^5$ and $10^6$ iterations.  In contrast, the HMC algorithm has already begun exploration of both modes in less than $10^2$ trajectories.  Not only that, but we also
observe little difference between the distributions at $10^4$ and $10^6$ trajectories, suggesting a very rapid convergence.  If we compare the distributions for both algorithms at $10^6$ iterations/trajectories, we see that the
HMC has converged to two modes of almost equal height, while the DEMC still suggests asymmetric modes.  To test the validity of the HMC result, we ran a $10^7$ iteration DEMC chain.  While still not fully converged, the 
marginalized posteriors for inclination were closer to equal height as predicted by the HMC algorithm.

Due to the high correlation with inclination, we see a similar picture with the posterior evolution for $D_L$.  For the DEMC chain, we see no signs of convergence, with the shape of the posterior changing at every
snapshot.  In contrast, the HMC chain is displaying clear signs of convergence, with the posteriors after $10^4$ and $10^6$ trajectories being almost identical.  As with the inclination, a comparison of both algorithms
at $10^6$ iterations/trajectories displays a significant difference in the shapes of the posteriors.

Next we treat the posterior distributions for $({\mathcal M}_c, \mu)$ together.  BNS1 is an almost equal mass binary with a mass ratio of $q = 1.0165$ (where we define $q = m_1 / m_2$).  This introduces a physical 
boundary in parameter space that the algorithms have to contend with.  In this case, as well as the predicted slow convergence, the DEMC displays signs that it is having trouble dealing with the barrier at
$m_1 = m_2$.  As a result, the posterior distribution is shifted away from the true value of $\mu$.  In contrast, the HMC seems to have no problem with this physical boundary, with the peak of the posterior
almost at the true values for both parameters.

Finally, we look at the sky angles.  As the SNR of the source is quite high ($\sim50$), we expect the sky resolution to be quite good.  Once again we see a difference in the speed of convergence between the two 
algorithms.  However, in this case, there is not so much difference between the posteriors at $10^4$ and $10^6$ iterations/trajectories in both cases.

\begin{figure*}
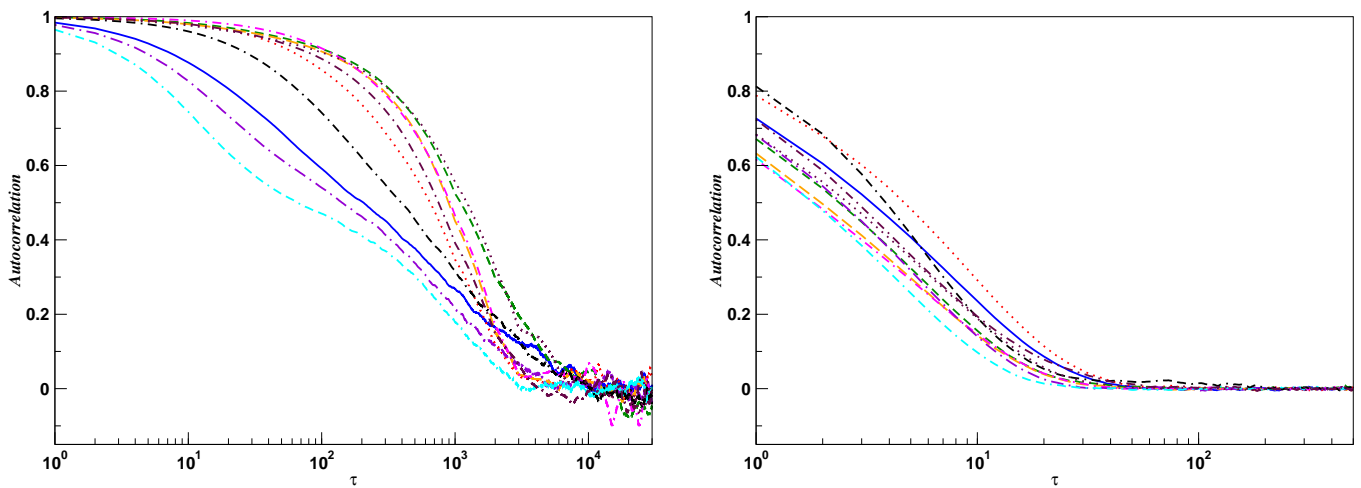

\begin{center}
\begin{minipage}[b]{.48\linewidth}
\centering
\epsfig{file=Article_All_BNS_DEMC_AutoCorr.eps, width=1.0\textwidth, height=2.5in}
\end{minipage} \hfill
\begin{minipage}[b]{.48\linewidth}
\centering
\epsfig{file=Article_All_BNS_HMC_AutoCorr.eps, width=1.0\textwidth, height=2.5in}
\end{minipage}
\caption{Autocorrelation of the slowest mixing chain as a function of lag $\tau$ for all BNS sources using a $10^6$ iteration DEMC chain (left panel) and a $10^{6}$ trajectory HMC chain (right panel).}
\label{fig:bns1_ac}
\end{center}
\end{figure*}

%This is the reason why we decided to use visual inspections by looking at various quantities. First, we investigated how the chains explore difficult parts of the parameter space connected with the multimodality of the posterior distribution. Then we compared the marginalised posterior distributions as inferred from different chain lengths to see if we can see visual signs of convergence. We also computed the median and $99\%$ credible intervals as inferred from the full $10^{6}$ iteration chain to look at the differences between the results from DEMC and HMC. Finally, we looked at the autocorrelation of the chains defined by

A useful tool for diagnosing MCMC algorithms is estimating the effective sample size ($ESS$), i.e. the number of statistically independent samples in a chain of length $N$.  The ESS is defined as
\begin{equation}
ESS = \frac{N}{\tau_{int}},
\end{equation}
where $\tau_{int}$ is the integrated autocorrelation time defined by
\begin{equation}
\tau_{int} = 1 + 2 \sum_{\tau=1}^{N-1} \rho (\tau),
\end{equation}
and $\rho(\tau)$ is the autocorrelation function at lag $\tau$, given by 
\begin{equation}
\rho(\tau)  = \dfrac{\displaystyle\sum_{i=1}^{N - \tau} \left( X_{i} - \overline{X} \right) \left( X_{i + \tau} - \overline{X} \right)  }{\displaystyle \sum_{i=1}^{N } \left( X_{i} - \overline{X} \right)^{2}}.
\end{equation}
Here $X_i$ denote the chain samples, and $\overline{X}$ is the sample mean.  If the chain is mixing well, we expect the autocorrelation to fall off to zero quite quickly, whereas a slowly mixing chain will have an 
autocorrelation that falls to zero at very high lags.  We should comment that when calculating $\tau_{int}$, noise in the autocorrelation at high lags can cause the integral to diverge, especially if the chain is 
not converged.  To ensure stability, we integrate $\tau_{int}$ to different values of $N$, and visually inspect the integral for convergence.

In Fig.~\ref{fig:bns1_ac}, we plot the autocorrelation for the slowest mixing chain, for each of the ten binary systems, using the DEMC (left panel) and HMC (right panel) algorithms.  In general we found that the
DEMC algorithm had autocorrelations that went to zero at lags of $3\times10^3 \leq \tau_{zac} \leq 1.4\times10^4$, where $\tau_{zac}$ is the zero autocorrelation lag.   We should point out that in all cases, the slowest
mixing chains were either for $\iota$ or $D_L$, confirming that the DEMC struggled with bi-modal exploration.  In contrast,  the HMC had autocorrelations that 
fell to zero almost two orders of magnitude faster, with zero-crossings at $50 \leq \tau_{zac} \leq 226$.  In terms of the $ESS$, the $10^6$ iteration DEMC chains produced sample sizes of between 
$283\leq ESS\leq 879$.  However, as expected, the HMC algorithm produces a much larger sample size with $47619\leq ESS \leq111111$, again an improvement of at least two orders of magnitude.  Full 
details can be found in Table \ref{tab:results_run_sources} of Appendix \ref{results_DEMC_HMC}.

While it is difficult to make an apples-to-apples comparison with the current algorithms in the LALInference package, we can compare the performance of the HMC algorithm with past studies conducted
by the LIGO/Virgo collaboration.  In Ref.~\cite{veitch_2014}, in a study using a Nested Sampling algorithm, and a similar TaylorF2 waveform model,  the CPU time to generate a statistically independent sample (SIS) 
is around $77.1$ seconds . In Ref~\cite{Berry_2014}, a similar study using a MCMC algorithm and the TaylorF2 waveform model, mentions that the total CPU time for a single run was $\sim280$ hours 
for an $ESS\sim 4500$, with an averaged CPU time to generate one SIS of $227$ seconds.  In general, the runtime for the HMC algorithm, using unoptimized waveform generation and likelihood calculation, was 24-28 hours. 
This translates to approximately one SIS per second.  The one exception was for BNS2.  In this case, the posterior distribution had a very complex geometry that required the generation of an additional 40000 hybrid/numerical
trajectories during Phase III.  However, even in this case, where the runtime was $\sim$106 hours, the algorithm still produced one SIS every 7 seconds.

Finally, in Table \ref{tab:Results_Table_Credible_Interval} of Appendix \ref{results_DEMC_HMC}, we quote the median values and $99\%$ credible intervals inferred from the DEMC and HMC chains for the parameters $\{m_{1}, m_{2}, m, q, D_{L},t_{c} \}$. While the value of the inclination is also of astrophysical interest, we omit it from the table due to its bi-modality. In all cases, we find that the true values are contained within the 99\%
credible intervals from the HMC chain.

\section{Conclusion}
In this work, we have designed a HMC algorithm for the parameter estimation of non-spinning BNS sources, using the ground-based network of detectors composed of Advanced LIGO/Advanced Virgo at design 
sensitivity.  The first step in this work was to define optimal values for the free parameters of the HMC algorithm, namely the mass matrix, as well as the step size and length of the Hamiltonian trajectory.   
A major computational cost for the HMC algorithm is the necessity of calculating gradients of the log-likelihood at multiple points along the Hamiltonian trajectories.  For GW astronomy, as there is no 
analytic closed-form solution to the log-likelihood, the gradients have to be calculated numerically.  The use of numerical gradients results in an unacceptable runtime and prohibits the use of the HMC
algorithm for real-time data analysis.  To circumvent the gradient calculation, we used an analytical fit for six of the nine parameters, and look-up tables for three problematic  parameters associated with bi-modal
posterior distributions.  This reduces the runtime, for a $10^6$ trajectory chain,  from many months to approximately one day, producing an effective sample size of (at least) many tens of thousands.

This work was a feasibility study that used a reduced parameter space for BNS parameter estimation.   Now that we believe the HMC to be a viable algorithm for ground-based Bayesian inference, work has 
already begun on extending this analysis to incorporating both spins and tidal deformation.  The gain in performances and computation time by using the HMC algorithm could first of all be of great 
importance for electromagnetic follow-up of GW sources and low-latency parameter estimation.

\bibliography{Article_HMC_v8.bib}

\onecolumngrid
\appendix

\section{BNS test-sources}
\label{BNS_table_sources}

In this work, we use a sample of BNS systems from a previous LIGO/Virgo parameter estimation study~\cite{Singer:2014qca}.  The sources were chosen to represent a range of mass ratios
and orientations.  In Table~\ref{tab:definition_sources_BNS} we provide the system parameters for each of the ten binaries, using the parameter definitions from the LIGO/Virgo study, i.e.
inclination $\iota$, phase at coalescence $\phi_c$, polarization angle $\psi$, right ascension $\alpha$ and declination $\delta$, all in degrees.  Individual masses $m_1$ and $m_2$ in solar
masses, luminosity distance $D_L$ in Mpc and time to coalescence $t_c$ in seconds.  In column 2 we provide the LIGO/Virgo catalog number as a point of reference.  In the final column,
we provide the three-detector signal-to-noise ratios.

\begin{table*}[t]
\begin{center}
\scriptsize
\begin{tabular}{|c|c|c|c|c|c|c|c|c|c|c|c|}
\hline
BNS & C\# & $\iota$/deg & $\phi_{c}$/deg & $\psi$/deg  & $D_L$/Mpc & $m_1 / M_{\odot}$ & $m_2 / M_{\odot}$ &$\alpha$ / deg & $\delta$ / deg  & $t_c$ / secs & $\rho_{HLV}$  \\
\hline
1 & 5384 & 46 & 105 & 315 & 43 & 1.23 & 1.21 & 216.4 & -77.7 & 31.92 & 49.01 \\
2 & 699 & 40 & 333 & 108 & 41 & 1.34 & 1.23 & 223.9 & 51.3 & 29.36 & 86.01 \\
3 & 899 & 26 & 139 & 118 & 84 & 1.36 & 1.25 & 99.9 & -30.8 & 28.58 & 51.77 \\
4 & 1135 & 149 & 162 & 342 & 57 & 1.43 & 1.24 & 168.9 & 9.0 & 27.61 & 34.04\\
5 & 1281 & 38 & 324 & 254 & 72 & 1.43 & 1.20 & 64.9 & 42.2 & 28.38 & 23.74\\
6 & 2608 & 153 & 305 & 215 & 46 & 1.36 & 1.35 & 106.1 & 16.7 & 26.80 & 64.95\\
7 & 2704 & 34 & 201 & 289 & 87 & 1.32 & 1.30 & 345.2 & 58.7 & 28.35 & 34.73\\
8 & 3015 & 176 & 327 & 115 & 68 & 1.43 & 1.30 & 277.7 & -19.8 & 26.53 & 50.05\\
9 & 3123 & 155 & 81 & 307 & 77 & 1.46 & 1.23 & 121.0 & 70.4 & 27.33 & 39.44\\
10 & 3249  & 145 & 110 & 141 & 83 & 1.31 & 1.31 & 77.8 & -25.9 & 28.35 & 50.28\\
\hline 
\end{tabular}
\caption{Source information for the BNS test-sources. The BNS number will be used as our source reference during this thesis, while $C\#$ is the reference number of each source in the 2 year EM follow-up study. The SNR, $\rho_{HLV}$, is the three detector network SNR retrieved using a template with the true parameter values.}
\label{tab:definition_sources_BNS}
\end{center}
\end{table*}

\section{Statistical Analysis}
\label{results_DEMC_HMC}

%\subsection{Computation time and run information}

In Table~\ref{tab:results_run_sources} we present a full runtime breakdown and statistical analysis of the HMC chains.  For each binary, columns 2-4 present the runtime (in hours)
for the initial 1500 numerical gradient trajectories in Phase I, the acceptance rate at the end of Phase I and the number of recorded data points during Phase I.  In general, this phase
takes between 2.5 and 5 hours, depending on the total mass, mass ratio and coalescence time.  Column 5 presents the time taken (in minutes)
to fit the $6\times220$ coefficients of the analytical gradient approximation, given by Eqn~(\ref{eq:gradient_approximation}), for the well behaved parameters.  As can be seen, this phase is very fast, with all
fits taking less than ten minutes.  In columns 6 and 7, we provide the runtime (in hours) for Phase III and the complete algorithm.  Except for BNS2, which required many more
hybrid/numerical trajectories, the Phase III runtime is on the order of a day.   We should highlight that using the various approximations for the gradients, 
the 985,000 trajectories in Phase III take between 5-6 times longer than the runtime for the 1500 numerical gradient trajectories in Phase I.

In the next two columns we give the number of extra hybrid and numerical trajectories required to 
fit the gradients.  Again, except for BNS2, less than $10^4$ extra fitting trajectories are needed.  BNS2, due to the complexity of the posterior distribution, required close to 
40,000 extra trajectories.  We are currently investigating situations like this to see if they can be optimized.  In column 10 we give the final acceptance rates for each of the HMC runs,  which are always $> 75\%$.  While not given here, in contrast, the DEMC algorithm had final acceptance rates of between 5\%-22\%.  Columns 11 and 12 detail the 
zero crossing lag for the autocorrelation function, and the integrated autocorrelation time.  In the final two columns, we present the effective sample size for both the HMC and DEMC
algorithms.  We can see that in all cases, the HMC produces approximately $10^4 - 10^5$ statistically independent samples.  This is always at least two orders of magnitude better
than what is produced with the DEMC algorithm.

\begin{table}[t]
\begin{center}
\scriptsize
\begin{tabular}{|c|c|c|c|c|c|c|c|c|c|c|c|c|c|}
\hline
BNS & $t_{PI}/hr$ & $AR_{PI}/\%$ & $N_{PI}$ & $t_{PII}/min$  & $t_{PIII}/hr$ & $t/hr$ & $T_{hyb}$ &$T_{num}$  & $AR/\%$  & $\tau_{zac}$ & $\tau_{int}$ & $ESS_{HMC}$  & $ESS_{DEMC}$\\
\hline
1 & 5.39 & 89.6 & 268800 & 8.4 & 22.2 & 27.8 & 599 & 55 & 85.8 & 63 & 17 &58823 & 396 \\
2 & 2.58 & 89.6 & 268800 & 6.3 & 102.8 & 105.5 & 37228 & 1007 & 86.0  & 146 & 19 & 52631 & 364\\
3 & 4.14 & 89.8 & 269400 & 5.2 & 21.9 & 26.2 & 3557 & 394 & 81.2  & 85 & 13 & 76923 & 283\\
4 & 4.03 & 93.1 & 279200 & 6.0 & 20.8 & 25.0 & 655 & 116 & 85.3 & 69 & 12 & 83333 & 410\\
5 & 2.52 & 94.1 & 282200 & 7.2 & 19.8 & 22.5 & 1104 & 100 & 83.3 & 51 & 18 & 55556 & 387\\
6 & 3.23 & 87.9 & 263600 & 6.6 & 21.9 & 26.3 & 3899 & 364 & 80.6 & 80 & 10 & 100000 & 737\\
7 & 4.34 & 90.5 & 271600 & 5.4 & 21.0 & 25.5 & 300 & 41 & 89.5 & 45 & 9 & 111111 & 879\\
8 & 2.74 & 91.7 & 275000 & 5.6 & 25.8 & 23.6 & 2854 & 387 & 80.7 & 226 & 21 & 47619 & 518\\
9 & 4.21 & 94.8 & 284400 & 6.1 & 23.8 & 28.1 & 7101 & 460 & 76.2 & 164 & 15 & 66667 & 601\\
10 & 3.76  & 90.3 & 271000 & 4.8 & 22.8 & 26.7 & 6060 & 440 & 77.0 & 68 & 12 & 83333 & 321\\
\hline 
\end{tabular}
\caption{Runtime and statistical properties for a $10^6$ trajectory HMC chain.  In columns 2-4 we give the runtime (in hours), acceptance rate and number of
recorded accepted data points using numerical gradient trajectories in Phase I.  Column 5 gives the time to fit the $6\times220$ coefficients of the analytic gradient approximation (in minutes).  Columns 6-7 give the Phase III and total runtimes (in hours).  In columns 8-9 we display the number of extra hybrid/numerical
trajectories needed during Phase III, with the final acceptance rate given in column 10.  In columns 11-13 we detail the lag at which the worst performing
chain goes to zero, the integrated autocorrelation time, and the number of statistically independent samples acquired from the chain.  As a comparison, in the 
final column, we provided the number of statistically independent samples acquired from a $10^6$ iteration DEMC chain.}
\label{tab:results_run_sources}
\end{center}
\end{table}

In Table~\ref{tab:Results_Table_Credible_Interval}, we plot the median values and 99\% credible intervals ($CI$) from both the HMC and DEMC chains for some of the most astrophysically interesting parameters.  To calculate
the credible intervals, we first investigate the skewness of the posterior distribution.  We take the distribution to be normally distributed if the sample skewness
\beq
b_1 = \frac{\sqrt{n(n-1)}}{n-2}\frac{m_3}{m_2^{3/2}},
\eeq
has values of $-0.25 \leq b_1 \leq 0.25$.  Here the second and third standardised moments $m_k$ are defined by
\begin{equation}
m_k = \sum_i \left(x_i - \bar{x} \right)^k / n,
\end{equation}
where $\bar{x}$ denotes the sample mean.  In this case, and defining the sample standard deviation for each of the different parameters as $\sigma_\m = (\sqrt{m_2})_\m$, we can write the symmetric 99\% $CI$s
\beq
CI = median(\lm)\pm 2.58\, \sigma_{\m}.
\eeq
If the sample skewness is outside of these bounds, we calculate the $CI$ by integrating over the marginalized posterior distribution, i.e.
\beq
\int_{CI} p\left(\lm | s\right) d\lm = 1-\alpha,
\eeq
where for a $99\%$ CI, $\alpha = 0.01$.  In each table element, we present the true value, the HMC result and the DEMC result respectively.
%\subsection{Median and credible intervals}

\begin{table}[t]
\begin{center}
\scriptsize
\begin{tabular}{|c||c|c|c|c|c|c|}
\hline
BNS number & $m_{1} / M_{\odot}$ & $m_{2} / M_{\odot}$ & $m / M_{\odot} $ & $q$ & $D_{L} / Mpc$ & $t_{c} / s$  \\
\hline
     &   1.23 & 	 1.21 &   	2.44 &  	1.017	& 43 &	31.91560 \\ 
1  & $1.291^{+0.079}_{-0.079}$	& $1.153^{+0.069}_{-0.069}$ &	 $2.4445^{+0.0221}_{-0.0047}$ &	$1.120^{+0.137}_{-0.137}$  &	$49.5^{+15.5}_{-25.4}$ &	$31.91550^{+0.00027}_{-0.00049}$ \\
 & $1.297^{+0.077}_{-0.077}$ & $1.148^{+0.067}_{-0.067}$ & $2.4452^{+0.0197}_{-0.0053}$ & $1.129^{+0.133}_{-0.133}$ & $48.1^{+16.1}_{-26.2}$ & $31.91548^{+0.00026}_{-0.00045}$ \\
\hline 
    & 1.34 & 1.23 & 2.57 & 1.089 & 41 & 29.32578 \\      
 2 & $1.352^{+0.067}_{-0.067}$ & $1.219^{+0.059}_{-0.059}$ &  $2.5713^{+0.0140}_{-0.0041}$ & $1.109^{+0.108}_{-0.108}$ & $38.2^{+22.2}_{-22.2}$ & $29.32575^{+0.00017}_{-0.00028}$ \\
 & $1.357^{+0.064}_{-0.064}$ & $1.215^{+0.064}_{-0.054}$ & $2.5719^{+0.0111}_{-0.0046}$ & $ 1.117^{+0.104}_{-0.104}$ & $34.4^{+23.4}_{-23.4}$ & $29.32574^{+0.00016}_{-0.00022}$ \\
 \hline
    & 1.36 & 1.25 & 2.61 & 1.088 & 84 & 28.58003 \\         
 3 & $1.392^{+0.089}_{-0.089}$ & $1.222^{+0.077}_{-0.077}$ & $2.6138^{+0.0217}_{-0.0065}$ & $1.13919^{+0.14441}_{-0.14441}$ & $72.4^{+25.5}_{-44.5}$ & $28.57997^{+0.00029}_{-0.00046}$ \\
    & $1.399^{+0.087}_{-0.087}$  & $1.216^{+0.074}_{-0.074}$ & $2.6148^{+0.0199}_{-0.0075}$ & $1.15021^{+0.14053}_{-0.14053}$ & $70.7^{+27.0}_{-41.3}$ & $28.57995^{+0.00027}_{-0.00044}$ \\
  \hline
    & 1.43 & 1.24 & 2.67 & 1.153 & 57 & 27.60872 \\
 4 &  $1.441^{+0.113}_{-0.113}$ & $1.231^{+0.095}_{-0.095}$ & $2.6717^{+0.0382}_{-0.0097}$ & $1.170^{+0.182}_{-0.182}$ & $52.5^{+20.0}_{-28.3}$ & $27.60868^{+0.00040}_{-0.00079}$ \\
& $1.448^{+0.110}_{-0.110}$ & $1.225^{+0.091}_{-0.091}$ & $2.6729^{+0.0372}_{-0.0109}$ & $1.181^{+0.176}_{-0.176}$ & $50.6^{+23.8}_{-23.8}$ & $27.60866^{+0.00041}_{-0.00076}$ \\
\hline
& 1.43 & 1.20 & 2.63 & 1.192 & 72 & 28.38391 \\          
5 & $1.443^{+0.138}_{-0.138}$ & $1.190^{+0.111}_{-0.111}$ & $2.6325^{+0.0492}_{-0.0144}$ & $1.213^{+0.228}_{-0.228}$ & $65.9^{+39.9}_{-39.9}$ & $28.38388^{+0.00072}_{-0.00112}$ \\
& $1.451^{+0.132}_{-0.132}$ & $1.183^{+0.116}_{-0.107}$ & $2.6342^{+0.0460}_{-0.0161}$ & $1.227^{+0.219}_{-0.219}$ & $63.4^{+39.9}_{-39.9}$ & $28.38385^{+0.00068}_{-0.00102}$ \\
\hline
 & 1.36 & 1.35 & 2.71 & 1.007 & 46 & 26.79953 \\
 6 & $1.427^{+0.078}_{-0.078}$ & $1.287^{+0.069}_{-0.069}$ & $2.7142^{+0.0183}_{-0.0043}$ & $1.109^{+0.121}_{-0.121}$ & $41.4^{+15.2}_{-15.2}$ & $26.79946^{+0.00021}_{-0.00036}$ \\
& $1.433^{+0.076}_{-0.076}$ & $1.282^{+0.067}_{-0.067}$ & $2.7149^{+0.0165}_{-0.0050}$ & $1.118^{+0.118}_{-0.118}$ & $40.2^{+15.4}_{-15.4}$ & $26.79945^{+0.00020}_{-0.00032}$ \\
\hline
 & 1.32 & 1.30 & 2.62 & 1.015 & 87 & 28.35058 \\
7 & $1.401^{+0.103}_{-0.103}$ & $1.225^{+0.088}_{-0.088}$ & $2.6268^{+0.0346}_{-0.0070}$ & $1.144^{+0.167}_{-0.167}$ & $85.3^{+31.4}_{-31.4}$ & $28.35043^{+0.00039}_{-0.00074}$ \\
& $1.409^{+0.100}_{-0.100}$ & $1.219^{+0.085}_{-0.085}$ & $ 2.6280^{+0.0295}_{-0.0082}$ & $1.156^{+0.162}_{-0.162}$ & $82.3^{+30.8}_{-30.8}$ & $28.35041^{+0.00036}_{-0.00063}$ \\
\hline
 & 1.43 & 1.30 & 2.73 & 1.100 & 68 & 26.53244 \\           
 8 & $1.460^{+0.096}_{-0.096}$ & $1.274^{+0.083}_{-0.083}$ & $2.7338^{+0.0229}_{-0.0074}$ & $1.146^{+0.150}_{-0.150}$ & $53.3^{+19.0}_{-25.8}$ & $26.53238^{+0.00031}_{-0.00048}$ \\
& $1.465^{+0.094}_{-0.094}$ & $1.269^{+0.080}_{-0.080}$ & $2.7346^{+0.0237}_{-0.0082}$ & $1.154^{+0.146}_{-0.146}$ & $52.3^{+19.2}_{-24.6}$ & $26.53237^{+0.00029}_{-0.00046}$ \\
\hline
 & 1.46 & 1.23 & 2.69 & 1.187 & 77 & 27.32872 \\        
 9 & $1.462^{+0.107}_{-0.107}$ & $1.228^{+0.102}_{-0.098}$ & $2.6904^{+0.0356}_{-0.0120}$ & $1.190^{+0.172}_{-0.172}$ & $61.3^{+36.5}_{-36.5}$ & $27.32872^{+0.00040}_{-0.00070}$ \\
 & $1.466^{+0.103}_{-0.103}$ & $1.225^{+0.104}_{-0.083}$ & $2.6912^{+0.0289}_{-0.0127}$ & $1.197^{+0.165}_{-0.165}$ & $60.8^{+35.6}_{-35.6}$ & $27.32871^{+0.00038}_{-0.00057}$ \\
 \hline
 & 1.31 & 1.31 & 2.62 & 1.000 & 83 & 28.34895 \\        
 10 & $1.386^{+0.085}_{-0.085}$ & $1.239^{+0.075}_{-0.075}$ & $2.6248^{+0.0208}_{-0.0049}$ & $1.118^{+0.137}_{-0.137}$ & $73.5^{+39.3}_{-39.3}$ & $28.34885^{+0.00027}_{-0.00045}$ \\
 & $1.392^{+0.085}_{-0.085}$ & $1.234^{+0.074}_{-0.074}$ & $2.6256^{+0.0213}_{-0.0057}$ & $1.129^{+0.136}_{-0.136}$ & $69.1^{+37.5}_{-37.5}$ & $28.34884^{+0.00025}_{-0.00042}$ \\
 \hline 
\end{tabular}
\caption{Median and $99\%$ credible intervals for the set of parameters $\{m_{1},m_{2},m,q,D_{L},t_{c}\}$ as inferred from the $10^{6}$ trajectory HMC and $10^{6}$ iterations DEMC chains. The first line is the true value for the source, the second line and third line are the results for the HMC and DEMC algorithms respectively.}
\label{tab:Results_Table_Credible_Interval}
\end{center}
\end{table}

\end{document}